\documentclass[11pt]{article}
\pdfoutput=1
\usepackage{jcappub}
\usepackage{amssymb,amsmath, graphicx}
\usepackage{latexsym}
\usepackage{mathrsfs}
\usepackage{hyperref}
\usepackage{verbatim}
\usepackage{bm}
\usepackage{mathtools}

\usepackage{mathtools}

\usepackage{xcolor}

\def\bea{\begin{eqnarray}}
\def\eea{\end{eqnarray}}
\def\be{\begin{equation}}
\def\ee{\end{equation}}
\def\ba{\begin{array}}
\def\ea{\end{array}}

\newcommand{\calH}{{\cal H}}

\def\d{{\rm d}}

\def\taur{{\tau_{\rm rec}}}

\def\0{{\boldsymbol 0}}

\def\k{{\boldsymbol{k}}}

\def\l{{\boldsymbol{l}}}

\begin{document}
\title{Signatures of  graviton masses on the CMB}
\author[a]{Philippe Brax,}
\author[b]{Sebastian Cespedes,}
\author[b]{and\ Anne-Christine Davis}
\affiliation[a]{Institut de Physique Th\'eorique, Universit\'e Paris-Saclay, CEA, CNRS, F-91191 Gif/Yvette Cedex, France}
\affiliation[b]{DAMTP, University of Cambridge, Wilberforce Road, Cambridge, CB3 0WA, UK}
\emailAdd{philippe.brax@ipht.fr}
\emailAdd{scespedes@damtp.cam.ac.uk}
\emailAdd{a.c.davis@damtp.cam.ac.uk}
\abstract{The  impact of the existence of gravitons with  non-vanishing masses  on the B modes of the Cosmic Microwave Background (CMB) is investigated. We also focus on putative modifications to the speed of the gravitational waves. We find that a change of the graviton speed shifts  the acoustic peaks of the CMB and then could be easily constrained. For the case of massive gravity, we show analytically how the B modes are sourced in a manner differing from the massless case   leading to a  plateau at low $l$ in the CMB spectrum. We also study the case when there are more than one graviton, and when pressure instabilities are present. The latter would occur in doubly coupled bigravity in the radiation era.  We focus on the case where a massless graviton becomes tachyonic in the radiation era whilst a massive one remains stable. As the unstable mode decouples from matter in the radiation era, we find that the effects of the instability is largely reduced on the spectrum of B-modes as long as the unstable graviton does not grow into the non-linear regime.
In all cases when both  massless and  massive gravitons are present, we find that the B-mode CMB spectrum is characterised by a low $l$ plateau together with a shifted position for the first few peaks compared to a purely massive graviton spectrum, a shift which depends on the mixing between the gravitons  in their coupling to matter and could serve as a hint in favour of the existence of multiple gravitons.  }
\maketitle
\section{Introduction}
A detection of primordial  gravitational waves would provide  new  insight into the inflationary epoch but would also  help  constraining non-standard features of gravity with  a higher degree of precision~\cite{Suzuki:2015zzg,Watts:2015eqa,Abazajian:2016yjj}. This would be the case if either the speed of gravitons or modifications of gravity lead to  non-vanishing  graviton masses. Given that the cosmic microwave anisotropy (CMB) is essentially a linear process all the intricacies, such as screening effects in the non-linear regime,  that  modifications of gravity possess would  be eluded.  Hence new  clues  about the nature of gravity might be obtained this way.

The photons of the CMB are polarised through Thomson scattering. This polarisation  leaves an imprint on both the   tensor and the scalar perturbations. Interestingly, the CMB polarisation    can be decomposed in terms of curless  and divergenceless components. As the curless component depends only on the initial power spectrum of tensors, it implies that those  B-modes when  projected on the sky  retain some information about the nature of gravitational waves  during both  matter and radiation dominated eras~\cite{Kamionkowski:1996zd, Zaldarriaga:1996xe}.

Even though General Relativity (GR)  has been passing all solar system tests and cosmological constraints with flying colours~\cite{Koyama:2015vza},  there might be  still room for modifications in new sectors, e.g. with gravitational waves. In particular,  it has proved  difficult  to constrain GR  at early times as   many of the interesting phenomenological properties  of  modified gravity models at low redsdhift  are  screened at very high redshifts~\cite{Clifton:2011jh, Joyce:2014kja}.

Experimentally, modifying gravity seems to be a plausible possibility. Even if it turned out that Einstein's GR were the actual theory of gravity, testing it by comparing it with alternatives is certainly worth pursuing and also would lead to a better understanding of GR's specificity. In their modern forms, modifications of gravity all involve screening mechanisms such that  most of their  effects are shielded  when considering dense sources. As gravitational waves are unscreened  for all cosmic times and scales  they may serve as sirens for modifications of gravity ~\cite{Jimenez:2015bwa,Brax:2015dma}

 In the last few  years new theories of massive gravity have emerged. For instance Lorentz breaking effects at early times could be introduced in a consistent way  and their cosmological implications have been studied~\cite{ArkaniHamed:2002sp}. For  Lorentz invariant massive gravity,  consistent theories  have been discovered fairly recently~\cite{deRham:2010ik, deRham:2010kj, Hassan:2011hr}. They may include more than one graviton, and matter may couple  differently to each metric~\cite{Hassan:2011zd}. Nevertheless a fully consistent cosmology has not been obtained yet mainly because of the presence of  stability issues~\cite{DAmico:2011eto, Gumrukcuoglu:2015nua, Enander:2014xga, Lagos:2015sya}.

On the other hand, there are simpler modifications of GR which could  also explain the late time acceleration of the expansion of the  universe. An important example is provided by the  Horndeski models~\cite{Deffayet:2009mn}.  These have the particularity of yielding second order equations of motions. On the gravitational side these theories modify the speed of the gravitational waves~\cite{Lombriser:2015sxa, Brax:2015dma, Bettoni:2016mij}, although they are very constrained now  at small redshifts by the  new gravitational wave measurements~\cite{TheLIGOScientific:2017qsa}, which  seem to imply that theories which modify the curvature terms of the Einstein equations are ruled out~\cite{GWHordensky}  as dark energy models.  This effect, which cannot be removed by  a disformal transformation, has important cosmological consequences and therefore can be constrained  using primordial gravitational waves detection \cite{Burrage:2016myt}. In that sense constraining the mass of the graviton seems to be an important goal to meet. Although observations of gravitational waves have so far put very strong limits~\cite{Raveri:2014eea, Cornish:2017jml}, they are at least 10 orders of magnitude above the cosmological relevant scales ~\cite{deRham:2016nuf}.

In this paper we will focus  on B-modes . These are the curl-free polarisation components, which only depend on the gravitational wave evolution from early times. In order to analyse the effects of a modified gravity sector we will assume that the initial conditions are adiabatic and that there is a detectable tensor to scalar ratio $r$. Thus all the important effects will be produced by a modification of gravity during  matter and radiation domination.
The presence of either a mass or a different speed from the speed of light for gravitons, and  the changes in the dispersion relation  leads to interesting effects on the low  $l$ B-modes.  For a large range of masses this yields  a characteristic plateau~\cite{Dubovsky:2009xk,Fasiello:2015csa} in the CMB spectrum. Since this effect cannot be produced by any other known effect, one may hope  to constrain gravity very precisely in this manner.

We have found an  analytical solution  showing that the source function of the B modes has a plateau until recombination, instead of being zero and then peaking at recombination as  happens in the massless case. This  plateau is then  projected onto the CMB power spectrum producing a plateau for the small  $l$ modes. Moreover the amplitude of the plateau  oscillates with the mass of the graviton, so this effect could be  used to constrain the mass of the graviton.

We also focus on the existence of multiple gravitons which could couple to matter with different strengths. This leads to a richer phenomenology but it also requires a more careful treatment.  We have been  able to  extend our analysis to models with more than one graviton and  will assume that there is no  hierarchy between the masses or the couplings of the gravitons to matter. We can diagonalise the coupled system of equations during matter and radiation  domination and thus  show how the signal behaves in different configurations. We  use this to show that even in the cases with more than one graviton the signal is still similar to that with one graviton, i.e. qualitatively different from the one in  massless gravity. The effect of the mixing between a massive and a massless graviton would also be characterised by a shift of the first few peaks of the B-mode CMB spectrum.

We  also study the effect of the instability problem of doubly coupled  bigravity  on the B-modes~\cite{Brax:2016ssf,Comelli:2015pua}. Indeed one of the gravitons becomes tachyonic in the radiation era. This  implies a power law behaviour as a function of the scale factor when the modes are out of the horizon and a possible  modification of the B mode spectrum. As the unstable mode does not couple to matter during pressure domination we find that its effect on the B-modes is reduced as long as the unstable mode remains in the perturbative regime. However, provided the initial conditions of the unstable mode are appropriately reduced~\cite{Comelli:2015pua,Brax:2017uwg}, its effect on the
first peaks of the B-mode spectrum is such that they can differ from the ones of a purely massive graviton. Hence if the low $l$ plateau characteristic of a massive graviton were observed, a shift in the position
of the first few peaks would be representative of the presence of another graviton mode. On the other hand,  these models also suffer from a gradient instability~\cite{Comelli:2015pua,Gumrukcuoglu:2015nua,Brax:2016ssf}, possibly lethal, in the vector sector whose study is left for future work.

In this paper we first  derive analytical solutions for massive gravity during  matter and radiation domination in section 2 and study the effects of a change of the speed of gravitons. We then calculate in section 3  how the  B-mode   power spectrum behaves for massive gravity, where we get an analytical solution at  low $l$.  We then examine  the effects of adding another graviton coupled to matter in section 4 and we include the case of a pressure instability in radiation domination.  Finally we conclude.

\section{General results}
\subsection{Massive graviton}
We want to investigate the propagation  of a  massive graviton, with mass $m$,  when the background cosmology is described by a  FRW (Friedmann-Robertson-Walker) Universe. We will focus  on the gravitational waves during matter and radiation domination, as these are the relevant ones for  the CMB. The graviton equation is \footnote{For a more thorough discussion see \cite{Bernard:2015mkk}.}
\bea
E''+\left(k^2+m^2a^2-\frac{a''}{a}\right)E=0,
\label{GW:equation}
\eea
where we have suppressed the indices and $E_{ij}= a h_{ij}$ where $h_{ij}$ represents the transverse and traceless part of the tensor perturbation.
Note that, $a''/a=(aH)^2+(aH)'$, where $aH$ is the size of the horizon. Then for $k^2+m^2a^2\ll \frac{a''}{a}$ modes are  out of the horizon  and evolve with constant amplitude $h_{ij}$ . The re-entry of the modes inside the horizon depends on the mass of the graviton now, contrary to the case of massless gravitons. In the following we will consider that the mass of the graviton is of order $H_0$ or larger. When $k^2+m^2a^2\gg \frac{a''}{a}$ the modes start oscillating with a frequency given by $\omega^2 \propto k^2+m^2a^2$ in the WKB approximation, which  leads to imprints on the B-mode spectrum. In order to investigate the precise nature of these oscillations we will solve (\ref{GW:equation}) during matter and radiation domination.
\subsubsection{Matter domination}
During matter domination  we have  that $a= \bar H_0^2 \tau^2 \propto H_0^2\tau^2$, where $\bar H_0 ={\cal O}(H_0)$  and  the graviton equation becomes
\bea
E''+\left(k^2+m^2\bar H_0^4\tau^4-\frac{2}{\tau^2}\right)E=0.
\eea
Let us  rewrite the equation in terms  $h_{ij}$  which reads
\bea
h''+\frac{4}{\tau}h'+(k^2+m^2\bar H_0^4\tau^4)h=0.
\eea
Using the variable $x=\frac{m\bar H_0^2\tau^3}{3}$, the wave equation becomes
\bea
\ddot h+\frac{2}{x}\dot h+(1+\frac{k^2}{m^2\bar H_0^4\tau^4})h=0,
\eea
where the dot means $d/dx$.

\begin{figure}[!ht]
\begin{center}
\includegraphics[scale=0.7]{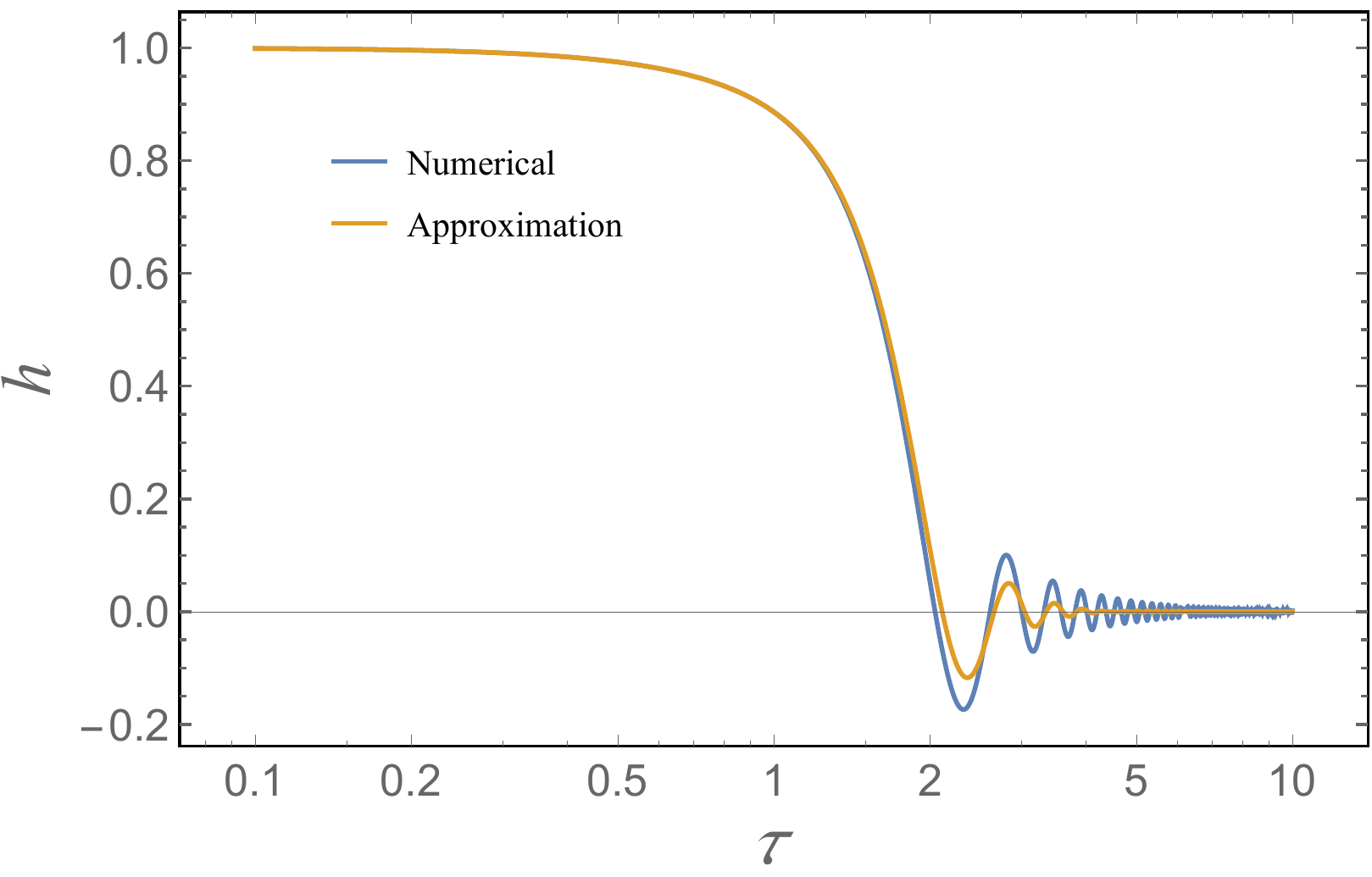}
\caption{The tensor mode solutions for $k=\bar H_0$ and $m=\bar H_0$ as function of conformal time in the matter era where the transition from constant to oscillatory behaviour is exemplified. The analytical approximation fits the numerical solution up to the onset of oscillations.  }
\label{matter:approx}
\end{center}
\end{figure}
Notice that  in the case where  $m^2a^2\gg k^2$, the equation simplifies and its solution is a  spherical Bessel function of order 0. This solution is only  constant at early times until  it enters the horizon and then decays. On the other hand when  $m^2a^2\ll k^2$, the field behaves as a   massless spin-2 field.
In order to obtain a solution which can be valid for a wide range of times  we note that the massive part modulates the massless part and we can approximate $h$ as
\bea
h=3\frac{j_1(k\tau)}{k\tau}\times j_0(\frac{1}{3}m\bar H_0^2\tau^3).
\label{massGrav:solution}
\eea
Because both solutions are constant at early times this combination has  the right behaviour there. The complete  solution  enters the horizon when $m^2\bar H_0^4\tau^6+k^2\tau^2=2$ and then oscillates. This is captured by the approximate solution as can be seen in
fig.\ref{matter:approx} where  the approximate solution is  very accurate up to horizon re-entry. On the other hand the amplitude of the oscillations within the horizon is suppressed due to the combined oscillatory behaviour of the two spherical Bessel functions at small scales. Here the approximate solution differs from the numerical one, but we still capture  the most important features for our analysis.
\subsubsection{Radiation domination}
To find the solutions of the wave equation  during  radiation domination we follow the same procedure as for the matter era. During this epoch $a= \hat H_0  \tau$  where $ \hat H_0 = \bar H_0^2 \tau_{\rm eq}$ where $\tau_{eq}$ is the conformal time at matter-radiation equality, the wave equation is now
\bea
h''+\frac{2}{\tau}h'+(k^2+m^2\hat H_0^2\tau^2)h=0.
\eea
Defining  $x=m\hat H_0\tau^2/2$ leads to
\bea
\ddot h+\frac{3}{2x}\dot h+\left(\frac{k^2}{2m\hat H_0x}+1\right)h=0.
\eea
When  $m^2a^2\gg k^2$, the solution is $\propto (m\hat H_0\tau^2/2)^{1/4}j_{-1/4}(m\hat H_0\tau^2/2)$ and when $m^2a^2\ll k^2$  it behaves like  $j_0(k\tau)$. A good solution  that can interpolate between both regimes is given by
\bea
h\propto (m\hat H_0\tau^2/2)^{1/4}j_{-1/4}(m\hat H_0\tau^2/2)j_0(k\tau).
\eea
where again we  have that the approximation is very accurate outside  the horizon but it oscillates too fast after re-entry.  The matching between the solutions in the radiation and matter eras  is presented in the Appendix.

\subsubsection{Superhorizon $k\tau\ll 1$ modes }
At wavelengths much larger than  the horizon the solution during  matter domination is more relevant, as the super horizon modes  enter later and make the most important contribution to the B-mode power spectrum.  We can expand the matter era solution as
\bea
h\sim \frac{3 \sin \left(\frac{m \bar H_0^2\tau ^3}{3}\right)}{m \bar H_0^2\tau ^3}\left(1-\frac{(k\tau)^2}{10}\right)+O\left(k^4\right).
\eea
Note that the first term goes to one when the mass is zero, and we recover the massless solution which is constant out of the horizon. As  $\tau$ grows, we can have two possibilities depending on whether  $\frac{m \tau ^3}{3}>k\tau$ or not. A given  mode enters the horizon earlier when the mass of the graviton is big enough to satisfy $\frac{m \tau ^3}{3}>k\tau$. In the contrary case the massive part will not lead to a significant modification of the  gravitational waves.

These new oscillation due to the mass  introduce an imprint on the B-modes which will  differ from the one coming from the massless modes.  As the contribution of the gravitational waves to the B-modes is dominated by the modes evaluated at $\tau=\taur$, for all wavenumbers  such that $\frac{m \taur ^3}{3}>k\taur$ the power spectrum  will be dominated by the  massive part of $h$.

\subsubsection{Modified tensor speed}
\begin{figure}[!ht]
\begin{center}
\includegraphics[scale=0.4]{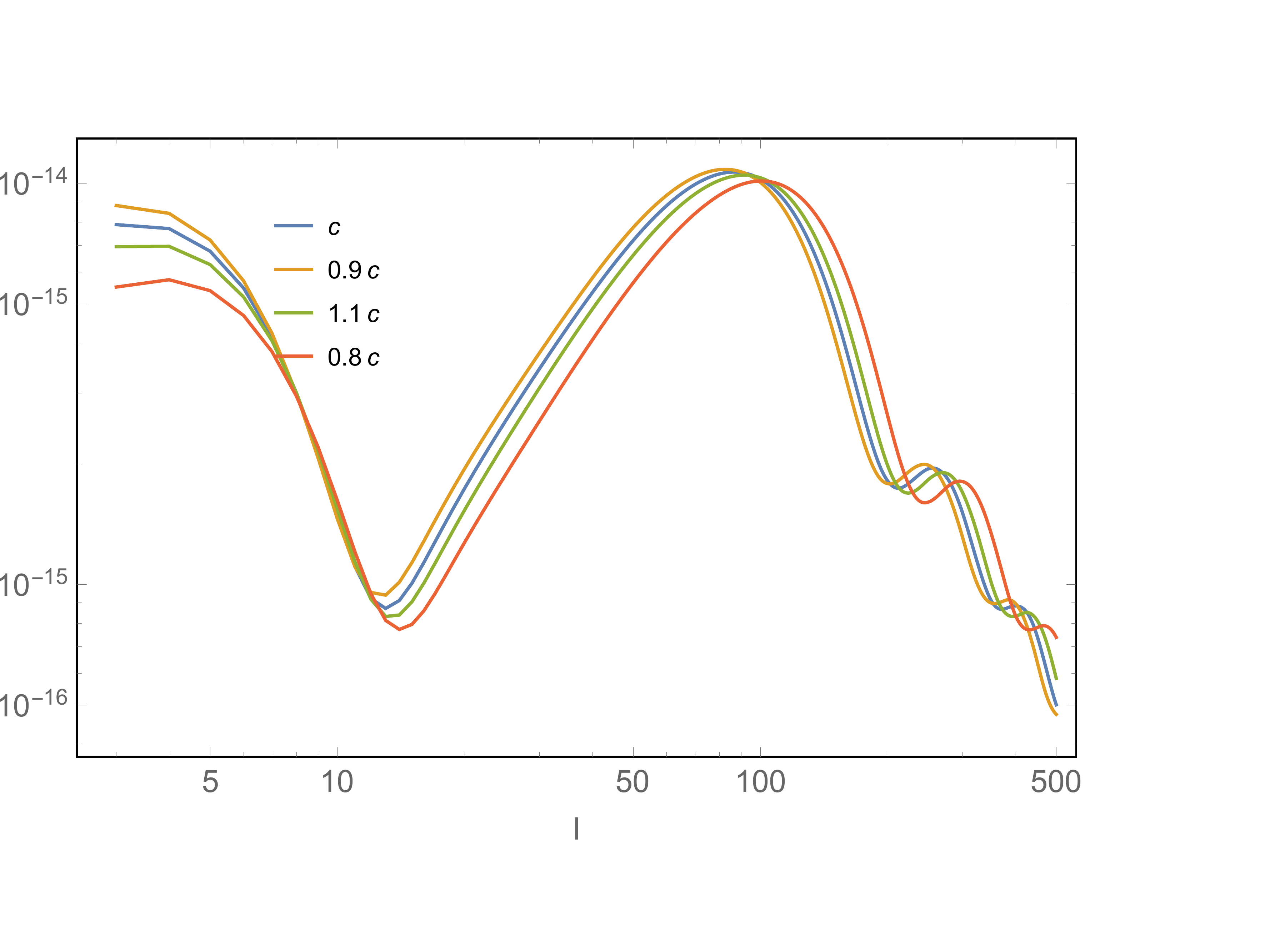}
\caption{CMB polarisation power spectrum for different values of the speed of the tensor modes $c_T$, in terms of the speed of  light $c$. The shift of the peaks due to a change of speed is apparent.  }
\label{modified:speed}
\end{center}
\end{figure}

We can also analyse the case when the modes propagate with a speed different from the speed of light. This scenario could arise in a variety of modifed gravity theories and only makes sense when the modification occurs in the early Universe as a very tight bound on the  deviation from the speed of light has been set by LIGO/VIRGO  in our local environment at small redshift~\cite{TheLIGOScientific:2017qsa}.  In this case the equation is simply,
\bea
h''+2aHh'+(c_T^2k^2)h=0.
\eea
The solution is the same as in   the massless case with a rescaled wavenumber. Hence  the gravitational wave will reach the horizon  at a shifted time. For example during  matter domination we have that
\bea
h''+\frac{4}{\tau}h'+(c_T^2k^2)h=0,
\eea
whose solution is  given by the spherical Bessel function
\bea
h=3\frac{j_1(c_T k\tau)}{c_Tk\tau}.
\eea
It is constant until $c_T k\tau=1$ after which  it decays to zero.
As the field  enters the horizon at a different time, it leaves a modified  signature on the CMB. All the acoustic peaks are shifted as  the source function, see below, is   rescaled by the same factor, which is shown in  figure \ref{modified:speed}.
Notice that  the location of the peaks are proportional to $l/c_T$, so by constraining their positions one could infer $c_T$. This is supposed to be at the sub-percent level for CMB stage 4 experiments \cite{Abazajian:2016yjj}, which is several orders of magnitude worse than the LIGO/VIRGO bounds.
One could try to analyse  the modification of the tensor speed by performing a disformal transformation, which changes the slope of the gravitational light-cone. When doing this the speed of the scalar part of the perturbations is  modified~\cite{Burrage:2016myt}. This would imply  that the acoustic peaks of the temperature power spectrum would be  shifted.

\section{Polarisation and  massive gravitons}
It is  instructive to rewrite the polarisation equations, and how they are modified in the presence of a massive  graviton and  a change in the speed of propagation of tensor modes.
We will focus only on B modes for now as they give  primordial information, although the results for E modes are similar.

Thomson scattering in the early universe generates a linear polarisation that can be best described by a $2\times 2$ traceless tensor involving  the $Q$ and $U$ Stokes parameters. It is convenient to pick up a particular combination of these two parameters which only depends on the tensor modes. These are called $B$ modes because they  have the parity of a magnetic field. The corresponding $E$ modes  also exist but they  depend  on the scalar modes and  will not be useful for our analysis.

The polarisation tensor state $\Psi$ can be expressed in term of temperature and polarisation multipoles as \cite{Zaldarriaga:1996xe},
\bea
\Psi&\equiv& \left[\frac{1}{10}\Delta_{T0}^{(T)}+\frac{1}{35}\Delta_{T2}^{(T)}+\frac{1}{210}\Delta_{T4}^{(T)}-\frac{3}{5}\Delta_{P0}^{(T)}+\frac{6}{35}\Delta_{P2}^{(T)}-\frac{1}{210}\Delta_{P4}^{(T)}\right].
\eea
The B modes power spectrum is given by
\bea
C_{Bl}=(4\pi)^2\int k^2\d k P_h(k)\left\vert\int_0^{\tau_0}\d\tau g(\tau)\Psi(k,\tau)\left(2j'_l(k\tau)+\frac{4j_l}{k\tau}\right)^2\right\vert,
\label{B-modes:pwspectr}
\eea
where $g(\tau)$ is the visibility function which is defined in terms of the optical depth for Thomson scattering $\kappa$ as
\bea
g(\tau)=\dot\kappa e^{-\kappa}.
\eea
 $P_h(k)\propto k^{n_T-3}$, with $n_T\sim0$ to leading order, is the primordial tensor power spectrum and the integral is between the initial time and now at $\tau_0$.
We denote by
\be
\Delta_{Bl}=\int_0^{\tau_0}\d\tau g(\tau)\Psi(k,\tau)\left(2j'_l(k\tau)+\frac{4j_l}{k\tau}\right)^2,
\ee
the source function.

\subsection{Boltzmann equations for tensor perturbations}
In order to calculate the temperature and polarisation  multipoles we need to analyse the Boltzmann equations associated with the Thomson scattering. These are~\cite{Zaldarriaga:1996xe},
\bea
\dot{\Delta}_T^{(T)}+ik\mu\Delta_T^{(T)}&=&-\dot h-\dot \kappa (\Delta_T^{(T)}-\Psi),\\
\dot{\Delta}_P^{(T)}+ik\mu \Delta_P^{(T)}&=&-\dot\kappa(\Delta_P^{(T)}+\Psi).\\
\eea
As an input we have that the metric perturbations evolve according to
\bea
\ddot{h}+2\calH\dot{h}+(k^2+m^2a^2)h=0.
\eea
For simplicity we will assume tight coupling. Then the equations can be expanded in terms of the visibility function $\dot\kappa^{-1}$. We then get,
\bea
\dot{\Delta}_{T0}&=&-\dot h -\dot\kappa[\Delta_{T0}-\Psi],\nonumber\\
\dot{\Delta}_{P0}&=&-\dot\kappa[\Delta_{P0}+\Psi],\nonumber\\
\dot{\Delta}_{Tl}&=&0, l\geq 1,\hspace{0.5cm}\dot{\Delta}_{Pl}=0, l\geq 1.
\label{Boltzmann:hierarchy}
\eea
These simplifications jointly with the definition of $\Psi=\frac{1}{10} \Delta_0^T$ lead to
\bea
\dot\Psi+\frac{9}{10}\dot\kappa\Psi=-\frac{\dot h}{10}.
\eea
This  gives
\bea
\Psi (\tau) =-\frac{1}{10}\int^{\tau}_0\d\tau' \dot h(\tau') e^{-\frac{9}{10}\left(\kappa(\tau')-\kappa(
\tau)\right)}.
\eea
Now we also assume that the visibility function is a peaked Gaussian distribution function during recombination, i.e.
\bea
g(\tau)\equiv\dot\kappa e^{-\kappa}=g(\taur)e^{-\frac{(\taur-\tau)^2}{\Delta\taur^2}}.
\eea
This
implies that during recombination we can approximate $\dot\kappa\approx\kappa/\Delta\taur$. Also assuming that $h$ varies slowly during recombination, we have then the approximation
\bea
\Psi(\tau) \sim -\frac{\dot h(\taur)}{10}e^{\frac{9}{10}\kappa(\tau)}\Delta\taur\int^{\infty}_1\frac{\d x}{x}e^{-\frac{9}{10}\kappa x},
\label{Psi:approximated}
\eea
where in the last integral we have introduced the  variable  $x=\kappa(\tau')/\kappa(
\tau)$. This changes the integration limits to 1 and $e^{\tau/\Delta\taur}$ which can be approximated to be infinity as long  $\Delta \taur$ is very small.

In the above calculation we have assumed that  $h(\tau)$ peaks around the value  $\taur$ as the visibility function is sharply peaked.
A better approximation  can be obtained  by averaging $h(\tau)$  through recombination. By assuming that the visibility function is a Gaussian as before we have
\bea
\dot h(\tau)\to \int^{\tau}_0 d\tau' g(\tau')\dot h(\tau')\sim \dot{h}(\taur)e^{-(k\Delta\taur)^2/2},
\label{doth_at_recombination}
\eea
for modes inside the horizon.

\subsection{Large angular scales}
\begin{figure}[!ht]
\begin{center}
\includegraphics[scale=0.7]{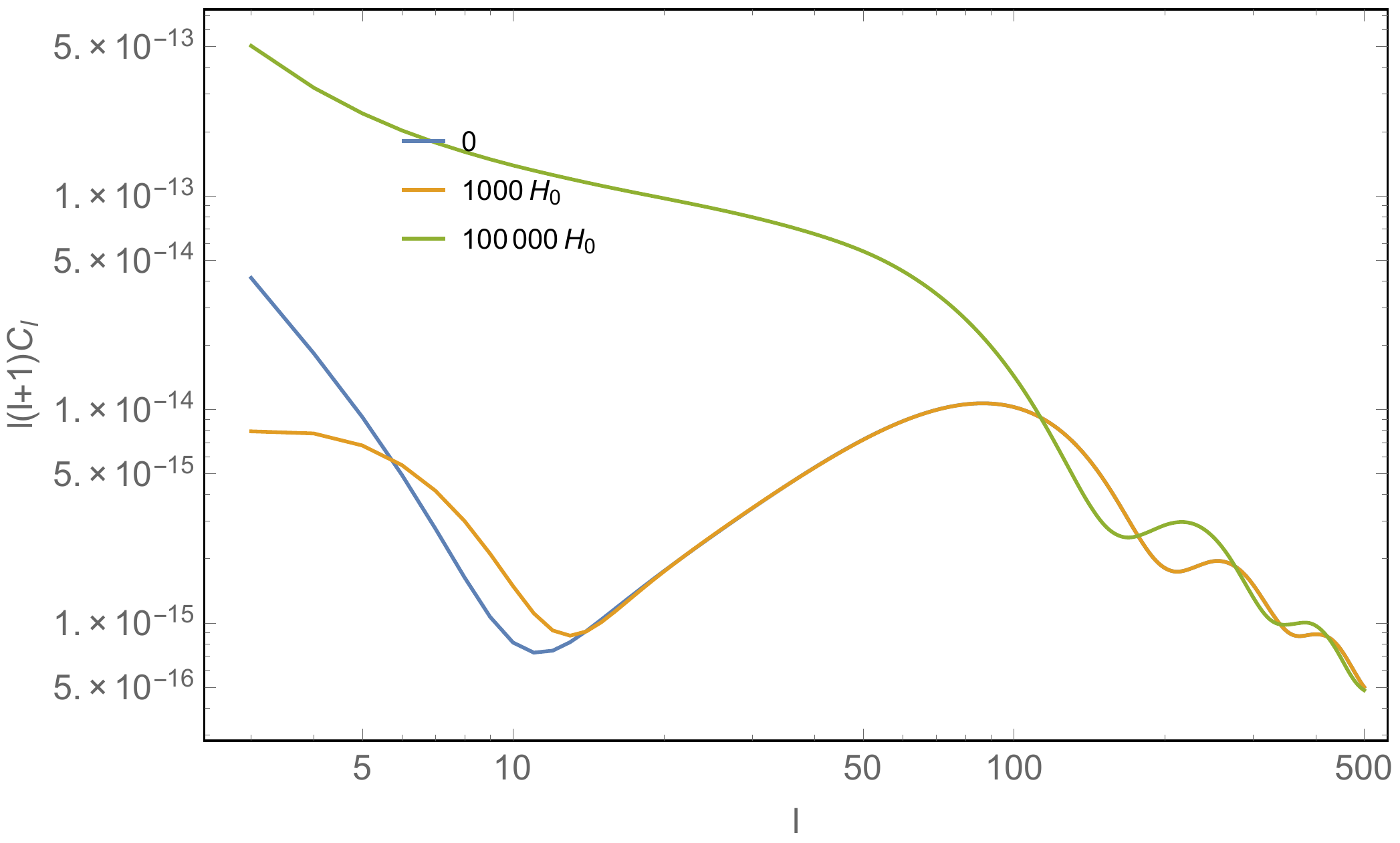}
\caption{The spectrum of B modes for different values of the graviton  mass.  }
\label{plot:massive}
\end{center}
\end{figure}

We have seen that there are important changes to the wave equation when the graviton is outside the horizon. As this effect is predominant at large angular scales we will proceed to study this sector in more detail. We first start with the massless case. Here, the solution of the graviton during matter domination is given by
\bea
h(\tau)=3\frac{j_1(k\tau)}{k\tau},
\label{graviton:solution}
\eea
where the factor of $3$ appears to normalise the wave function  to $1$ at $k\tau=0$. This solution is constant initially and then  decays in an oscillating fashion after $k\tau=1$ . We are interested in modes that enter the horizon around  the time of recombination. These modes correspond to scales which  are so large that when they  enter  the horizon the universe is in matter domination. Now on such  large scales the effects of recombination are not relevant so we can approximate  $g(\tau)=\delta(\tau-\taur)$ . The spectrum (\ref{B-modes:pwspectr}) of B-modes becomes
\bea
C_{BB,l}^T\propto
\int k^2\d k P_h(k)\left\vert\dot h(k\taur)\left[2j_l'(k(\tau_0-\taur))+4\frac{j_l(k(\tau_0-\taur))}{k(\tau_0-\taur)}\right]\right\vert^2.
\eea
In the massless case the graviton wave function is constant until it enters the horizon.  As the behaviour of the integral is dominated by $k\approx l/(\tau_0-\taur)$, we have the approximation
\bea
C_{BB}^{l}\approx(k^5 P_h(k,\taur))\vert_{k\approx l/(\tau_0-\taur)}\int \d\ln x\left\vert 2j'_l(x)+\frac{4j_l(x)}{x}\right\vert^2,
\eea
where $\dot h (k\taur) \propto k\taur$ on large scales.
The integral of the spherical Bessel function scales as  $\propto l^{-2}$ for small $l$,  then we have that  for large scales,
\bea
l(l+1)C_{BB}^l\sim l(l+1),
\eea
which grows linearly with $l$ for small $l$. In the massive case, the large scale behaviour is very different, see figure \ref{plot:massive} for instance.
\subsubsection*{Massive graviton case}
In the case of a massive graviton during matter domination the solution is approximately   given by
\bea
h=3\frac{j_1(k\tau)}{k\tau}\times j_0(\frac{1}{3}m\bar H_0^2\tau^3).
\eea
We will focus on the  case where $ma\gg k\approx l/(\tau_0-\taur)$ as  the B-mode spectrum  will be stronger at small $l$, and we comment on the other cases. In this regime the graviton has  a constant amplitude until it enters the horizon. After entering the horizon  the graviton will behave as if massless.
On  large angular scales the main difference with the massless case springs from  the source function.
Assuming a Gaussian visibility function, and that projection factors involving spherical Bessel functions  vary slowly over the last scattering surface, using (\ref{Psi:approximated}) and (\ref{doth_at_recombination}), the source function  becomes
\bea
\Delta_{Bl}(k)&=&P_{Bl}[k(\tau_0-\taur)] \int_0^{\tau_0}\d\tau g(\tau)\psi(\tau)\nonumber\\
&=&P_{Bl}[k(\tau_0-\taur)]\frac{\dot h(\taur)}{10}\Delta\taur e^{-(k\Delta\taur)^2/2}\int_0^{\infty}\d\kappa e^{-\kappa}\int^\infty_1\frac{\d x}{x}e^{-\frac{9}{10}\kappa x},
\eea
where
\bea
P_{Bl}[k(\tau_0-\taur)]=\vert 2j_l'(k(\tau_0-\taur))+4\frac{j_l(k(\tau_0-\taur))}{k(\tau_0-\taur)}\vert^2,
\eea
which   implies that the source function  can be written as,
\bea
\Delta_{Bl}(k)=P_{Bl}[k(\tau_0-\taur)]\dot h(\taur)\Delta\taur e^{-(k\Delta\taur)^2/2}\left(\frac{1}{10}\log\frac{19}{9}\right).
\eea
As we are considering  the regime where $m^2 \bar H_0^4\taur^4\gg k^2$ we can use the approximate solution (\ref{massGrav:solution}) to calculate $\dot h(\taur) $,
\bea
\dot h(\taur)\approx \frac{\cos(\frac{m\bar H_0^2\taur^3}{3})}{\taur}(1-\frac{(k\taur)^2}{10}).
\label{LargeScalesh}
\eea
Now we  can calculate the power spectrum of B-modes at large scales.  As the source function falls off inside the horizon, the integral over $k$ is dominated by large scales corresponding to the low $l$ modes. In this case the spherical Bessel functions are  constant at low $l$.  Therefore the B-mode spectrum on large scales is essentially sensitive to (\ref{LargeScalesh}) yielding
\bea
C_{l}^B\propto \frac{\cos^2(\frac{m\bar H_0\taur^3}{3})}{\taur^2}\propto \frac{1}{2\taur^2}\left(1+\cos\left(\frac{2m\bar H_0\taur^3}{3}\right)\right).
\label{Approx:CLLB}
\eea
which is a constant amplitude corresponding to a plateau whose amplitude oscillates with the mass $m$.

We can evaluate  the maximal angular multipole $l$ where effects of a massive graviton can be observed. The dispersion relation of a massive graviton changes from the massless case   when $k\approx m\times a$ then at recombination this corresponds to
\bea
l\approx k(\tau_0-\taur)&=&m\times a(\taur)\times(\tau_0-\taur)\nonumber\\
&\approx & \frac{m}{H_0}(1+z_r)^{-1}\int^{1}_{(1+z_R)^{-1}}\frac{\d x}{\sqrt{\Omega_\Lambda x^4+\Omega_m x+\Omega_r}}\approx 3.3(1+z_r^{-1})\frac{m}{H_0}\nonumber\\
\eea
where for $z_R\approx 1080$ we have that the B-modes are not modified  for masses below $300 H_0$.

\subsection{Flat sky limit}
In order to calculate the $l$ dependence of  the B-modes angular power spectrum we use  the flat sky approximation, which implies that we Fourier expand instead of projecting using spherical harmonics. In the flat sky limit the  amplitude of the scalar and polarisation modes   can be written as,
\bea
a_T^{\zeta}&=&\int_0^{\tau_0}\int\d\tau\frac{\d^3k}{(2\pi)^3}\zeta(\k)e^{-ik^zD}S_T^\zeta(\k,\tau)(2\pi)^2\delta^{(2)}(\k^{\parallel}D-\l),\\
a_B^{h}&=&\int_0^{\tau_0}\int\d\tau\frac{\d^3k}{(2\pi)^3}\sum_{\pm}\pm h^{\pm}e^{-ik^zD}(2i\frac{k^z}{k})S_P^h(\k,\tau)(2\pi)^2\delta^{(2)}(\k^{\parallel}D-\l),
\eea
where $S^X$ are the source functions which in the case of B modes is $-g\Psi(k,\tau)$ and in the case of scalar modes is given by  the sum of the two scalar gravitational potentials. Also $D=\tau-\tau_0$. This will greatly simplify calculations and we can see for example that the $\langle BB\rangle$ correlation is given by
\bea
\langle a_B^{h}(\l)a_B^h(\l')\rangle=\int\frac{\d^3k_1}{(2\pi)^3}\frac{\d^3k_2}{(2\pi)^3}2\langle hh\rangle\Delta_B^h(k_1^z,l_1)\Delta_B^h(k_2^z,l_2)(-4)\frac{k_1^z}{k_1}\frac{k_2^z}{k_2}(2\pi)^4\delta^{(3)}
(\k_1+\k_2),\nonumber\\
\eea
where we have defined the transfer function as,
\bea
\Delta_B^h(k^z,l)=\int_0^{\tau_0}\d\tau S_B^h(k)e^{-ik_zD}\delta^{(2)}\left(\k^{\parallel} D-\l\right).
\eea
 We first note that by reducing the product of delta functions we have
\bea
\langle a_B^{h}(\l)a_B^h(\l')\rangle&=&\int\frac{\d^3k_1}{(2\pi)^3}\frac{\d^3k_2}{(2\pi)^3}2\langle hh\rangle\Delta_B^h(k_1^z,l_1)\Delta_B^h(k_2^z,l_2)(-4)\frac{k_1^z}{k_1}\frac{k_2^z}{k_2}\delta^{(3)}
(\k_1+\k_2)\nonumber\\
&=&\int\frac{\d k^z\d \k_1^{\parallel} \d \k_2^{\parallel}}{(2\pi)^2}\int_0^{\tau_0}\d\tau\int_0^{\tau_0}\d\tau'8\langle hh\rangle\frac{(k^z)^2}{k_1k_2} S_B^h(k_1)e^{-ik_zD} S_B^h(k_2)e^{+ik_zD'}\nonumber\\
&&\times\delta^{(2)}\left(\k^{\parallel}_1 D-\l\right)\delta^{(2)}\left(\k^{\parallel}_2 D'-\l'\right)\delta^{(2)}
(\k^\parallel_1+\k^\parallel_2),\nonumber\\
\eea
where we have written $\d k^3=\d k^z\d \k^{\parallel}$. The integral is dominated by $D\sim D'$ and can be rewritten as
\bea
\langle a_B^{h}(\l)a_B^h(\l')\rangle&=&
\int\frac{\d k^z}{(2\pi)^2}\vert \int_0^{\tau_0}\frac{\d\tau}{D}\sqrt{8\langle hh}\rangle\frac{(k^z)}{k} S_B^h(k)e^{-ik_zD}\vert ^2\times\delta^{(2)}
(\l+\l')\nonumber\\
&=&\int\frac{\d k^z}{(2\pi)^2}\vert \int_0^{\tau_0}\frac{\d\tau}{D}\sqrt{8\langle hh}\rangle\frac{(k^z)}{\sqrt{(k^z)^2+l^2/D^2}} S_B^h(\sqrt{(k^z)^2+l^2/D^2})e^{-ik_zD}\vert ^2\times\delta^{(2)}(\l+\l').\nonumber\\
\label{Bmodes:reduced}
\eea
Now to evaluate the integral over time we will use the fact that the visibility function is a Gaussian distribution, which implies that
approximately
\bea
\kappa=\kappa_r e^{-(\tau-\taur)/\Delta\taur}.
\eea
We now define the integrand
\bea
\tilde{\Delta}(k,l)&=&\int_0^{\tau_0}\frac{\d\tau}{D}\sqrt{8\langle hh}\rangle\frac{(k^z)}{\sqrt{(k^z)^2+l^2/D^2}} S_B^h(\sqrt{(k^z)^2+l^2/D^2})e^{-ik_zD}\nonumber\\
&=&\int_0^{\tau_0}\frac{\d\tau}{D}\frac{\sqrt{8\langle hh}\rangle(k^z)}{\sqrt{(k^z)^2+l^2/D^2}} \frac{\dot h(\taur)}{10} e^{-(\sqrt{(k^z)^2+l^2/D^2}\Delta\taur)^2/2}\nonumber\\
&&\times g(\taur)e^{-(\tau-\taur)/\Delta\taur}e^{-\kappa}\int_1^{\infty}\frac{\d x}{x}e^{-\frac{9}{10}\kappa x}e^{-ik_zD},
\eea
where we have replaced $S^h_B$ by its value  from (\ref{Psi:approximated}). Notice  that the integral  gets most of its contribution around $\taur$ because of the exponential  function of the opacity which vanishes before recombination. Then $e^{-\kappa}\int_1^{\infty}\frac{\d x}{x}e^{-\frac{9}{10}\kappa x}=e^{-\kappa}\Gamma(0, \frac{9}{10}\kappa)$, where $\Gamma(a,b)=\int_b^{\infty}t^{a-1}e^{-t}\d t$ is the incomplete Gamma function.  The integral  decays for large values of $\kappa$ and is order one for $\kappa\sim\kappa_r$ so we  can neglect it in the  final expression. Furthermore,   we can replace $D\sim \Delta\taur$.
Then we get
\bea
\tilde{\Delta}(k,l)&\sim&\frac{\sqrt{8\langle hh}\rangle(k^z)}{\sqrt{(k^z)^2+l^2/D_r^2}} \frac{g(\taur)\dot h(\taur)}{10} e^{-(\sqrt{(k^z)^2+l^2/D_r})^2\Delta\taur^2/2}e^{-ik_zD_r}.\nonumber
\eea
Replacing this into (\ref{Bmodes:reduced}), we have
\bea
\langle a_B^{h}(\l)a_B^h(\l')\rangle&=&
\int\frac{\d k^z}{(2\pi)^2}\frac{8\langle hh\rangle(k^z)^2}{(k^z)^2+l^2/D_r^2} \frac{g(\taur)^2\dot h^2(\taur)}{100} e^{-\left((k^z)^2+l^2/D_r^2\right)\Delta\taur^2}e^{-2ik_zD_r}
\delta^{(2)}(\l+\l').\nonumber\\
\eea
We can approximate this integral by the stationary phase method. The saddle point is $k^z\approx -i\frac{D_r}{\Delta\taur^2}$ implying that   the power spectrum becomes
\bea
\langle a_B^{h}(\l)a_B^h(\l')\rangle&=&\frac{A_sr}{4\pi^{3/2}}\frac{g(
\taur)^2\dot h^2(\taur)}{100}e^{-\frac{D^2}{\Delta\taur^2}-\frac{ l^2\Delta\taur^2}{D_r^2}}\delta^{(2)}(\l+\l').\nonumber\\
&=&\frac{A_sr}{4\pi^{3/2}}\frac{g(
\taur) ^2\cos^2\left(\frac{m \bar H_0^2 \taur^3}{3}\right)}{100\taur^2}e^{-\frac{D^2}{\Delta\taur^2}-\frac{ l^2\Delta\taur^2}{D_r^2}}\delta^{(2)}(\l+\l').
\eea
where  we have replaced $\langle hh\rangle=Ark^{-3}$ with $A$ the amplitude of the primordial scalar perturbations and $r$ the tensor to scalar ratio and in the last line we have used (\ref{LargeScalesh}).

This result implies that  the modes will stay constant until $l\approx \frac{D_r}{\Delta \taur} \approx 100$. Although this result is very simplified it shows that the oscillations due to the graviton mass  modifies strongly the power spectrum. Note that the damping effect is independent of the mass of the graviton, see figure \ref{plot:massive} for instance. 
\section{Bigravity}
\subsection{Propagating modes}
In this section we consider the case of bigravity \cite{Comelli:2015pua, Brax:2016ssf,Gumrukcuoglu:2015nua} where one graviton is massless and the other is massive in a Minkowski background. This is inspired
by the bigravity case of massive gravity although we only use the gravitational sector of this model and do not deal with problems such as the range of validity of the model related to the existence
of strong coupling issues and vectorial instabilities in the radiation era. In our case, we only use this model as an illustration for the type of physics induced by the presence of two gravitons.
The CMB signal in this case  involves the two gravitons
implying that
\bea
\Psi\approx\left[\beta_1\times \dot h_1+\beta_2\times \dot h_2\right]^2,
\eea
where $\beta_{1,2}$ are coupling constants.
In the generic case the gravitational waves  propagate in different FRW metrics characterised two scale factors $a_{1,2}$ and the ratio between the two lapse functions $b$ leading to the two wave equations in vacuum
\bea
E_1''+k^2E_1+(M_{11}^2a_1^2-\frac{a_1''}{a_1})E_1+M_{12}^2a_1a_2 E_2&=&0,\nonumber\\
E_2''+b^2k^2E_2+(M_{22}^2a_2^2-\frac{a_2''}{a_2})E_2+M_{21}^2a_1a_2 E_1&=&0,
\label{doublecoupled:equations}
\eea
where we have dropped the tensorial indices so $E_1$ should be understood as $E^1_{ij}$, and $M_{12}=M_{21}$. For instance in the case of doubly coupled bigravity we have
\bea
b=\frac{a_1H_1}{a_2H_2},
\eea
for one of the two branches of cosmological backgrounds. Moreover $b\equiv 1$ in both matter and radiation eras. Here the conformal time is such that the Hubble rates are defined by
\be
H_{1,2}=\frac{da_{1,2}}{a_{1,2}^2 d\eta},
\ee
implying that when $b=1$ the two scale factors are proportional with $\beta_1 a_2= \beta_2 a_1$ \cite{Brax:2016ssf}.
Then during matter domination since $a_{i}= \frac{\beta_i}{\beta_1^2 +\beta_2^2} H_0^2 \tau^2$ such that $a= \beta_1 a_1 +\beta_2 a_2= H_0^2 \tau^2$,  we have that the above equations  become
\bea
E_1''+\left(k^2+M_{11}^2\bar H_0^4\tau^4-\frac{2}{\tau^2}\right)E_1+M_{12}^2\bar H_0^4\tau^4 E_2&=&0,\nonumber\\
E_2''+\left(b^2k^2+M_{22}^2\bar H_0^4\tau^4-\frac{2}{\tau^2}\right)E_2+M_{21}^2\bar H_0^4 \tau^4E_1&=&0,
\eea
where we have redefined $M^2_{ij}\to \frac{\beta_i\beta_j}{(\beta_1^2 +\beta_2^2)^2} M^2_{ij}$.
We want to find a particular combination of the two gravitons which satisfies  an equation for a massive graviton
\bea
f''+\left(k^2+M_f^2\bar H_0^4 \tau^4-\frac{2}{\tau^2}\right)f=0,
\eea
where $f=\lambda_1 E_1+\lambda_2 E_2$ and the coefficients $\lambda_1$ and $\lambda_2$ are constant. This implies that
\bea
\lambda_1(M_f^2-M_{11}^2)-\lambda_2 M_{12}^2=0,\nonumber\\
\lambda_2\left(\frac{k^2}{\tau^4}(1-b^2)+M_f^2 -M_{22}^2\right)-\lambda_1 M_{21}^2=0,
\label{sols:diag}
\eea
which admits non-trivial solutions for
 $M_f$ given by
\bea
2M_f^2=(b^2-1)\frac{k^2}{\tau^4}+M_{11}^2+M_{22}^2\pm\sqrt{4M_{12}^2M_{21}^2+\left((b^2-1)\frac{k^2}{\tau^4}+M_{11}^2-M_{22}^2\right)^2}.
\eea
In the following we assume that  $b=1$ corresponding to a single speed for the two gravitons.
We can rewrite the expression for  $M_f$ as,
\bea
2M_f^2=M_{11}^2+M_{22}^2\pm\sqrt{\Delta+(M_{11}^2+M_{22})^2},
\eea
where  $\Delta=4M_{12}^2M_{21}^2-4M_{11}^2M_{22}^2$. There are no tachyonic instabilities when  $\Delta$  takes values between  $0$ and $-(M_{11}^2+M_{22})^2$. In the latter one of the modes is  massless and we have that, $M_{11}^2M_{22}^2=M_{12}^2M_{21}^2$, which implies that  the other mode has a mass
\begin{figure}[!ht]
\begin{center}
\includegraphics[scale=0.7]{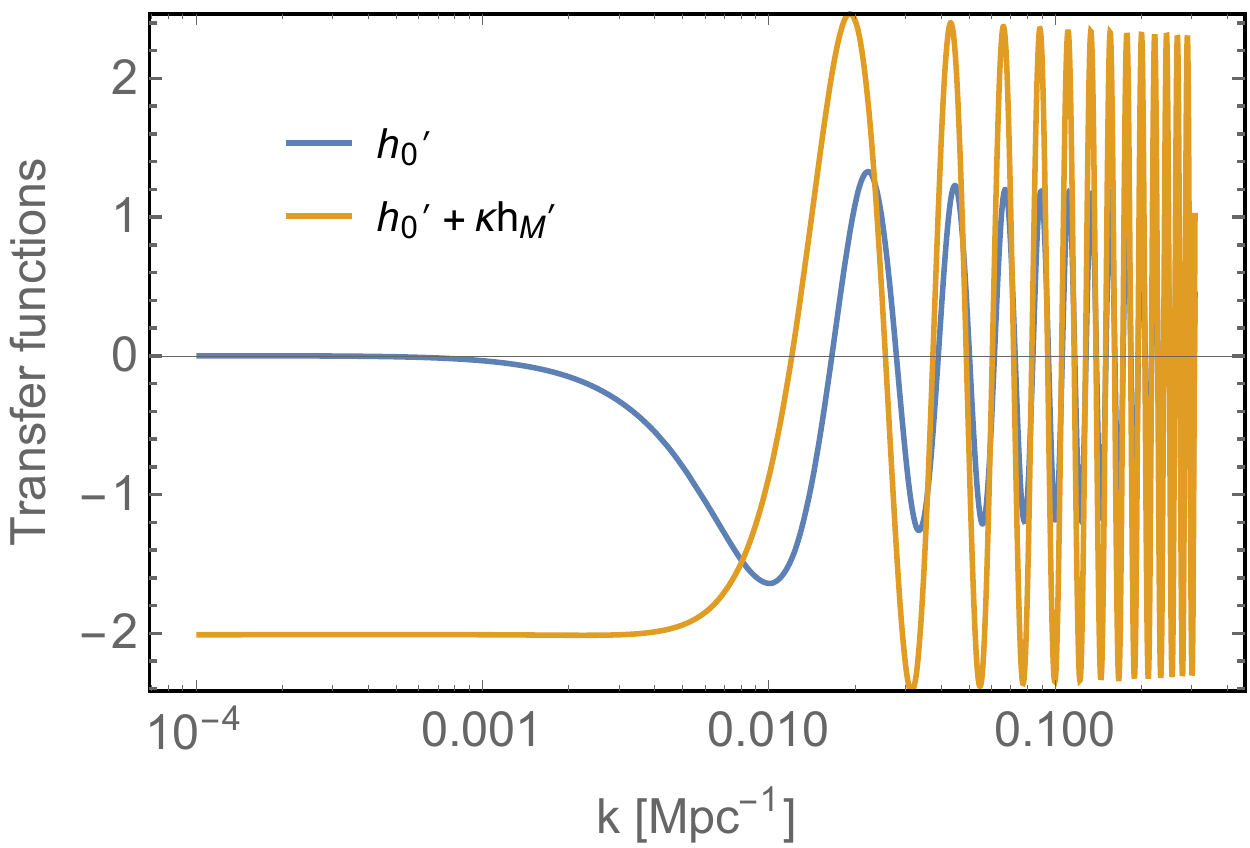}
\caption{Gravitational coupling $\Psi$ (\ref{Psi:approximated}), for the massless case and a combination of a massive and a massless graviton with $\kappa=0.1$ and the mass of the graviton is $1000H_0$. Notice that
$\Psi$ does not vanish on large scales in the massive case.  }
\label{BigravityTransfer}
\end{center}
\end{figure}

\bea
M_f^2=M_{11}^2+M_{22}^2.
\label{diagonalised:mass}
\eea
In the generic case, there is  always a light and a more massive mode.

\subsection{Coupling to matter}
 The impact of two propagating gravitons on the CMB B-mode spectrum  depends on how they source  the polarisation terms.
The coupling to matter is of the form
\bea
\delta S_{\rm{matter}}=\frac{1}{2m_{\rm Pl}}\int \d^4 x a_J\left\{\beta_1E^{(1)}_{ij}+\beta_2E^{(2)}_{ij}\right\}T^{ij},
\eea
where $a_J=\beta_1a_1+\beta_2a_2\equiv a $ is the Jordan frame scale factor, i.e. the scale factor of the FRW metric which couples to matter. We have also introduced the Planck scale $m_{\rm Pl}$ for the
normalised gravitons $E_{1,2}$. In the following we focus on the case with one massless graviton after diagonalisation corresponding to the constraint $\Delta=0$. When diagonalising the two graviton equations, the eigenmodes are
\bea
f_0=\alpha_1E_1+\alpha_2 E_2\nonumber\\
f_m=\alpha_3E_1+\alpha_4 E_2
\eea
where $f_0$ is the massless mode in the matter era, and $f_m$ the massive one. Imposing that the gravitons remain normalised for the   eigenmodes of the mass matrix implies that
the change of basis is a two dimensional rotation. Indeed we must have  $(\partial f_0)^2+(\partial f_m)^2=(\partial E_1)^2+(\partial E_2)^2$, therefore
\bea
\alpha_1&=&\cos\theta,\hspace{1cm}\alpha_2=- \sin\theta,\nonumber\\
\alpha_3&=&\sin\theta,\hspace{1cm}\alpha_4= \cos\theta.
\eea
Now using  the fact that $\Delta=0$ implies the existence of a massless graviton and a massive one of squared mass $M_{11}^2+M_{22}^2$  and replacing these values of the two masses into (\ref{sols:diag}) we obtain that,
\bea
\alpha_1&=&-\frac{M_{12}^2}{M_{11}^2}\alpha_2\\
\alpha_3&=&\frac{M_{12}^2}{M_{22}^2}\alpha_4.
\eea
As a result  we have that
\bea
E_1=\frac{1}{\sqrt{1+\tan^2 \theta}}(f_0+\tan\theta f_m)\\
E_2=\frac{1}{\sqrt{1+\tan^2 \theta}}(f_m-\tan\theta f_0)
\label{soll}
\eea
where
\bea
\tan\theta=\frac{M_{11}^2}{M_{12}^2}=\frac{M_{12}^2}{M_{22}^2}.
\eea
Then in the Jordan frame the gravitons are coupled to matter as
\bea
\delta S_{\rm{matter}}=\frac{1}{2m_{\rm Pl}\sqrt{1+\frac{M_{12}^4}{M_{22}^4}}}\int \d^4 x a\lambda\left\{\beta_1\left(f^0_{ij}+\frac{M_{11}^2}{M_{12}^2} f^m_{ij}\right)+\beta_2\left(f^m_{ij}-\frac{M_{12}^2}{M_{22}^2}f^0_{ij}\right)\right\}T^{ij},\nonumber\\
\label{JordaFrame:coupledTomatter}
\eea
which implies that,
\bea
\delta S_{\rm{matter}}=\frac{1}{2m_{\rm Pl}\sqrt{1+\frac{M_{12}^4}{M_{22}^4}}}\int \d^4 xa^2\beta_2\left\{\left(\frac{\beta_1}{\beta_2}-\frac{M_{12}^2}{M_{22}^2}\right)\frac{f^0_{ij}}{a}+\left(1+\frac{\beta_1}{\beta_2}\frac{M_{12}^2}{M_{22}^2}\right)\frac{f^m_{ij}}{a}\right\}T^{ij},\nonumber\\
\eea
where we have reintroduced the tensorial indices.
Now in bigravity the physical Planck scale is given $M_{\rm Pl}= \frac{m_{\rm Pl}}{\sqrt{ \beta_1^2 +\beta_2^2}}$ \cite{Brax:2016ssf} so that the coupling to matter becomes
\bea
\delta S_{\rm{matter}}=\frac{1}{2M_{\rm Pl}}\int \d^4x \frac{1}{\sqrt{1+\frac{M_{12}^4}{M_{22}^4}}\sqrt{ \beta_1^2 +\beta_2^2}}a\beta_2\left\{\left(\frac{\beta_1}{\beta_2}-\frac{M_{12}^2}{M_{22}^2}\right)\frac{f^0_{ij}}{a}+\left(1+\frac{\beta_1}{\beta_2}\frac{M_{12}^2}{M_{22}^2}\right)\frac{f^m_{ij}}{a}\right\}T^{ij}.\nonumber\\
\label{Diagonalised:couplings}
\eea
It is easy to see that the power spectrum of the graviton coupled to matter goes to the one of a single graviton coupled to matter at the end of inflation. Indeed taking that both $f_0/a$ and $f_m/a$ go to one at the end
of inflation, and upon using the statistical independent of $f_0$ and $f_m$, the power spectrum of the coupled graviton ${\cal P}_h$ is
\be
{\cal P}_h = \frac{\beta_2^2}{({1+\frac{M_{12}^4}{M_{22}^4}})({ \beta_1^2 +\beta_2^2})}\left (\left(\frac{\beta_1}{\beta_2}-\frac{M_{12}^2}{M_{22}^2}\right)^2 {\cal P}_0 + \left(1+\frac{\beta_1}{\beta_2}\frac{M_{12}^2}{M_{22}^2}\right)^2{\cal P}_m\right )
\ee
where ${\cal P}_{0,m}$ are the power spectra of the massless and massive gravitons.
Outside the horizon we normalise the massless and massive spectra similarly implying that
\be
{\cal P}_h\equiv{\cal P}_0={\cal P}_m
\ee
and therefore  the spectrum of the coupled graviton obtained from (\ref{Diagonalised:couplings}) is automatically normalised in the same fashion as in General Relativity. To analyse the effect of the coupling we introduce the parameter $\kappa$ as
\bea
\kappa=\frac{1+\frac{\beta_1}{\beta_2}\frac{M_{12}^2}{M_{22}^2}}{\frac{\beta_1}{\beta_2}-\frac{M_{12}^2}{M_{22}^2}}.
\eea
The gravitational source  becomes of the form $\Psi\propto f_0'+\kappa f_m'$. In  figure (\ref{BigravityTransfer}) we have plotted the power spectrum produced by (\ref{Diagonalised:couplings}) for different absolute values of the coupling $\kappa$. Notice that the effects of the massive graviton cannot be removed unless $\beta_2/\beta_1 <0$, which would lead to instabilities, or if there is no coupling $M_{12}=0$. In the generic case, the
characteristic plateau of massive graviton at low values of $l$ is always  present. Moreover the position of the first peaks is shifted when the coupling $\kappa$ varies.

\begin{figure}[!ht]
\begin{center}
\includegraphics[scale=0.5]{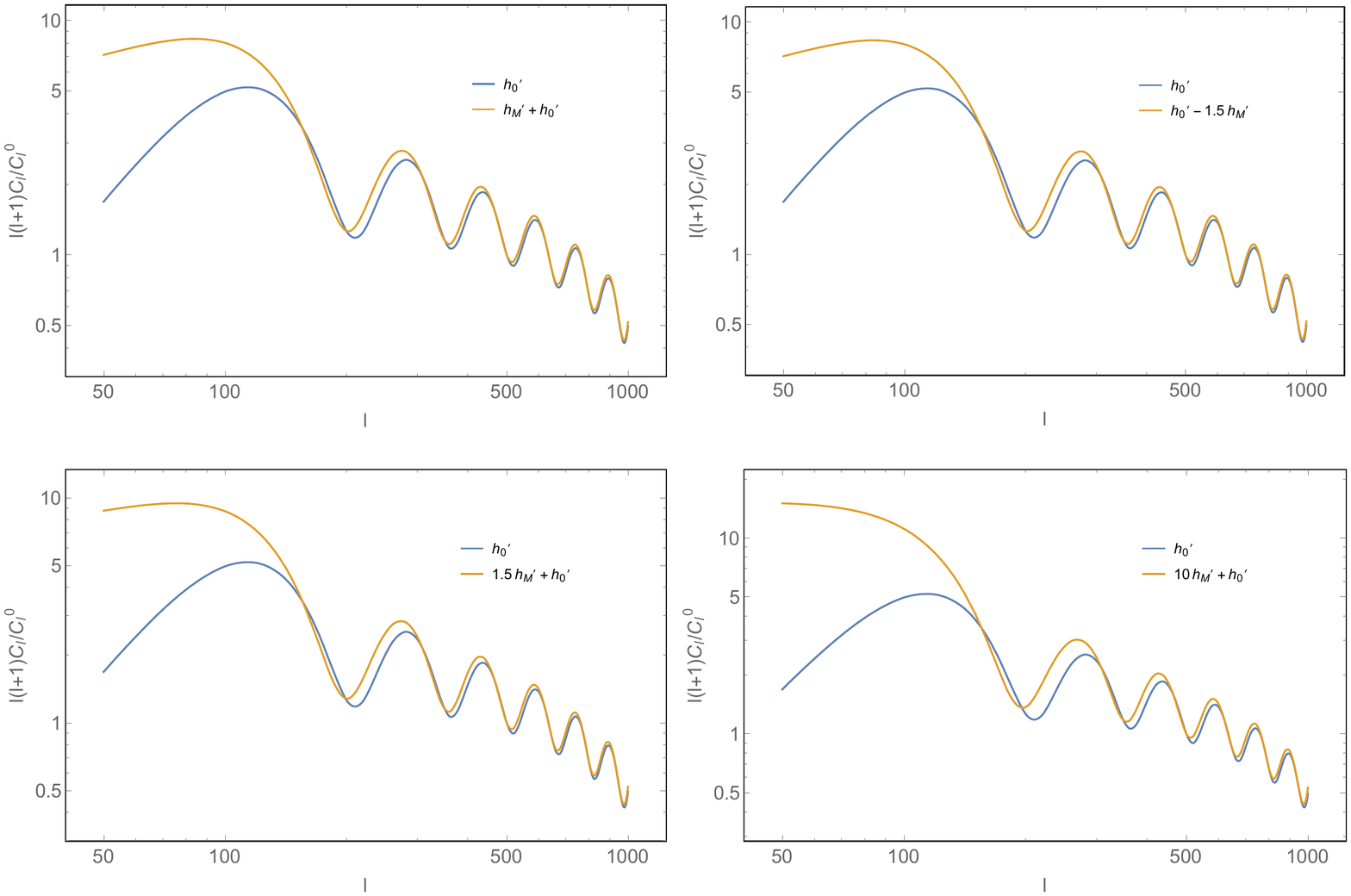}
\caption{ The power spectrum $C_{ll}^B$ for different values of $\kappa$ in the case of two gravitons, one being massive whilst the other one is massless. The mass is taken to be $1000H_0$. Notice that the plateau at low $l$ is always present and that the first few peaks shifts with the value of $\kappa$}
\label{BigravityTransfer}
\end{center}
\end{figure}

\subsection{Instabilities}
In doubly coupled bigravity models there is an  instability  which  appears in the radiation era coming from pressure terms such that  ~\cite{Brax:2016ssf}
\bea
\delta S_p=\frac{1}{8}\int\d^4x\sqrt{-g}\delta T_{ij}\delta g^{ij}.
\eea
It turns out that this yields  a pressure-dependent   mass matrix of the form,
\begin{figure}[!ht]
\begin{center}
\includegraphics[scale=0.45]{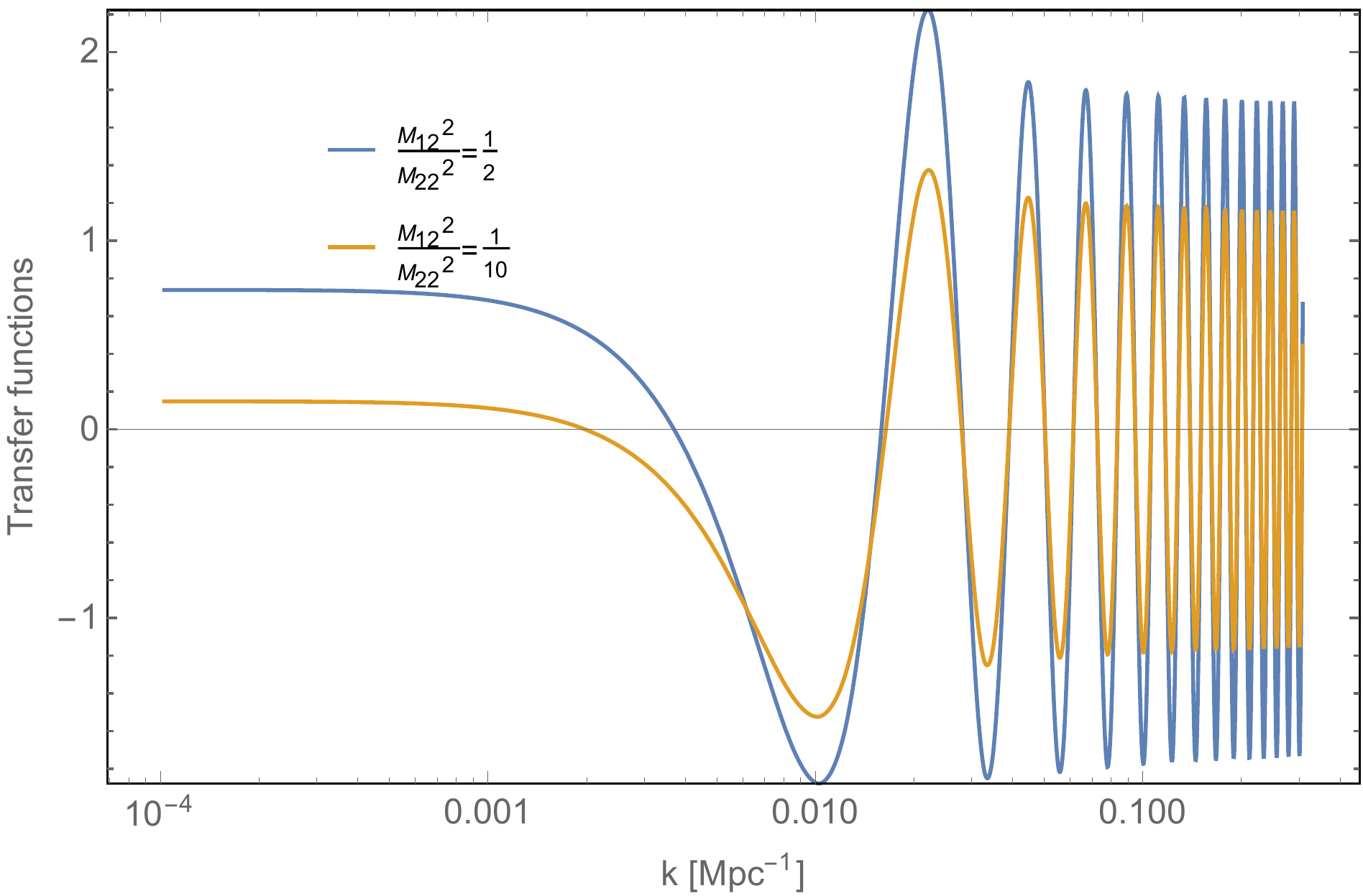}
\caption{The gravitational coupling $\Psi$   for a massive graviton together with an unstable massless one. We have taken the mass to be $1000 \ H_0$ and the ratio
$M^2_{12}/M^2_{22}$ to be $1/10$ and $1/2$ respectively. The amplitude for the unstable mode is multiplied by $10^{-11}$, i.e. this corresponds to the very low initial conditions of the unstable mode
which preserves the perturbativity of the model and gives a non-negligible effect on the B-mode spectrum.  }
\label{CMB_instab1}
\end{center}
\end{figure}
\bea
\Delta M_p^2=\frac{3\omega a_j^2H_j^2}{\beta_1^2+\beta_2^2}
\left(\begin{array}{cc}
-\beta_2^2 &\beta_1\beta_2\\
\beta_1\beta_2 & -\beta_1^2
\label{instability:massmatrix}
\end{array}\right),
\eea
which  has a zero mass eigenstate and an eigenmode of negative mass squared, i.e. a tachyon,
\bea
m_G^2=-3\omega_j^2a_j^2H_j^2<0.
\eea
This is only present during radiation domination where it  produces a mild instability. To analyse its effect we can solve the equation for a massive graviton during radiation domination
\bea
h''+2aH h'+(k^2+m^2a^2-a^2H^2)h=0.
\eea
where  we have that $a\propto \tau$ and  the equation reduces to
\bea
h''+\frac{2}{\tau} h'+(k^2+m^2\hat H_0^2 \tau^2-\frac{1}{\tau^2})h=0,
\eea
whose solution can be approximated by,
\bea
h\propto (m\hat H_0\tau^2/2)^{1/4}j_{-1/4}(m\hat H_0\tau^2/2)\left(j_0(k\tau)+j_{1/2 (-1 + \sqrt{5}))}(k\tau)\right).
\label{instability:solution}
\eea
The new spherical Bessel mode function $j_{1/2 (-1 + \sqrt{5}))}(k\tau)$ arises because of the instability in the radiation era.
This can be matched to the matter era using
\bea
h=
\begin{cases}
(m\hat H_0\tau^2/2)^{1/4}j_{-1/4}(m\hat H_0\tau^2/2)\left(j_0(k\tau)+j_{1/2 (-1 + \sqrt{5}))}(k\tau)\right) & \tau<\tau_{eq}\\
 3\frac{1}{k\tau}\times j_0(\frac{1}{3}m\bar H_0^2\tau^3)\left(Aj_1(k\tau)+By_1(k\tau)\right) & \tau>\tau_{eq}
\end{cases},
\eea
where $A$ and $B$ could be found by matching the solution and its derivative at $h(\tau_{eq})$. For other more realistic approaches see~\cite{Pritchard:2004qp}, where the WKB approximation is  used.
This new solution, i.e. its effect on the gravitational coupling $\Psi$,  is plotted in fig.\ref{CMB_instab1}. { We see that the amplitude of the modes is very large for small $k$  and that this contribution is of similar shape as for  the stable mode,  albeit with  a much higher amplitude. As a result,  the power spectrum for these modes has an amplitude of several orders of magnitude higher than for the stable case, which has to be compensated by the choice of very low initial amplitudes for the unstable mode \cite{Comelli:2015pua, Brax:2017uwg}. }

Now to analyse the  effect of the instability  in the case of two gravitons  we need to include (\ref{instability:massmatrix}) in (\ref{doublecoupled:equations}) and then diagonalise the new set of equations. We will do this  in radiation domination, and then we will match to matter domination. The equations are,
\bea
E_1''+\left(k^2+M_{11}^2\hat H_0^2\tau^2-\frac{\beta_2^2}{\beta_1^2+\beta_2^2}\frac{1}{\tau^2}\right)E_1+\left(M_{12}^2\hat H_0^2 \tau^2+\frac{\beta_1\beta_2}{\beta_1^2+\beta_2^2}\frac{1}{\tau^2}\right)E_2&=&0,\nonumber\\
E_2''+\left(k^2+M_{22}^2\hat H_0^2\tau^2-\frac{\beta_1^2}{\beta_1^2+\beta_2^2}\frac{1}{\tau^2}\right)E_2+\left(M_{21}^2\hat H_0^2 \tau^2+\frac{\beta_1\beta_2}{\beta_1^2+\beta_2^2}\frac{1}{\tau^2}\right)E1&=&0.
\label{Instability:equations}
\eea
As before we try to find solutions of the form $f=\lambda_1 E_1+\lambda_2 E_2$ where the equation for $f$ satisfies
\bea
f''+\left(k^2+M_f^2\hat H_0^2 \tau^2-\frac{1}{\tau^2}\right)f=0.
\label{inst}
\eea
Let us define $\tilde{M_{ij}}^2=M_{ij}^2+\frac{\beta_i\beta_j}{\beta_1^2+\beta_2^2}\frac{1}{\hat H_0^2\tau^4}$ in which case  we have that
\bea
\lambda_1({M_f}^2-\tilde M_{11}^2)-\lambda_2\tilde M_{12}^2=0,\nonumber\\
\lambda_2\left( M_f^2 -\tilde M_{22}^2\right)-\lambda_1\tilde M_{12}^2=0,
\eea
which is similar  to the case treated previously without an instability.
The  expression for $M_f$ reduces to,
\bea
2M_f^2&=&
\tilde M^2_{11}+ \tilde M_{22}^2 \pm\sqrt{(\tilde M^2_{11}+\tilde M_{22}^2)^2 +4(\tilde M_{12}^4- \tilde M_{11}^2 M_{22}^2)}.
\eea
In the following we focus on the case where $M_{12}^4= M_{11}^2 M_{22}^2$. 
Notice that deep in the radiation era, as was already  the case in the matter era, we have two solutions corresponding to a massless graviton $M_f=0$ which becomes unstable in the radiation era
and a massive graviton of mass $M_f^2= M_{11}^2 + M_{22}^2$ in both the radiation and matter eras. 

Similarly we can use the results (\ref{soll})  to diagonalise
\bea
 E_1=\frac{1}{\sqrt{1+\frac{\tilde M_{11}^4}{\tilde M_{12}^4}}}(f_0+\frac{\tilde M_{11}^2}{\tilde M_{12}^2} f_m),\\
 E_2=\frac{1}{\sqrt{1+\frac{\tilde M_{11}^4}{\tilde M_{12}^4}}}(f_m-\frac{\tilde M_{12}^2}{\tilde M_{22}^2})f_0
\eea
where we have denoted by $f_0$ the mode with $M_f=0$ and $f_m$ the massive one.
Notice that the diagonalisation is only valid when the rotation matrix is time independent, i.e. at all times  apart from the transitory regime where both $M_{ij}^2$ and $\frac{\beta_i\beta_j}{\beta_1^2+\beta_2^2}\frac{1}{\hat H_0^2\tau^4}$ are of the same order. In the following, we will neglect this intermediate regime
as we are either interested in the early times where the instability is present or later when it has disappeared.

We can again use the results from the previous sections and then  analogously to (\ref{Diagonalised:couplings}) we have that the coupling to matter reads
\bea
\delta S_{\rm{matter}}=\frac{1}{2M_{\rm Pl}}\int \d^4x \frac{1}{\sqrt{1+\frac{\tilde M_{12}^4}{\tilde M_{22}^4}}\sqrt{ \beta_1^2 +\beta_2^2}}a\beta_2\left\{\left(\frac{\beta_1}{\beta_2}-\frac{\tilde M_{12}^2}{\tilde M_{22}^2}\right)\frac{f^0_{ij}}{a}+\left(1+\frac{\beta_1}{\beta_2}\frac{\tilde M_{12}^2}{\tilde M_{22}^2}\right)\frac{f^m_{ij}}{a}\right\}T^{ij}.\nonumber\\
\label{InstabilityDiagonalised:couplings}
\eea
which is also naturally normalised.

\begin{figure}[!ht]
\begin{center}
\includegraphics[scale=0.65]{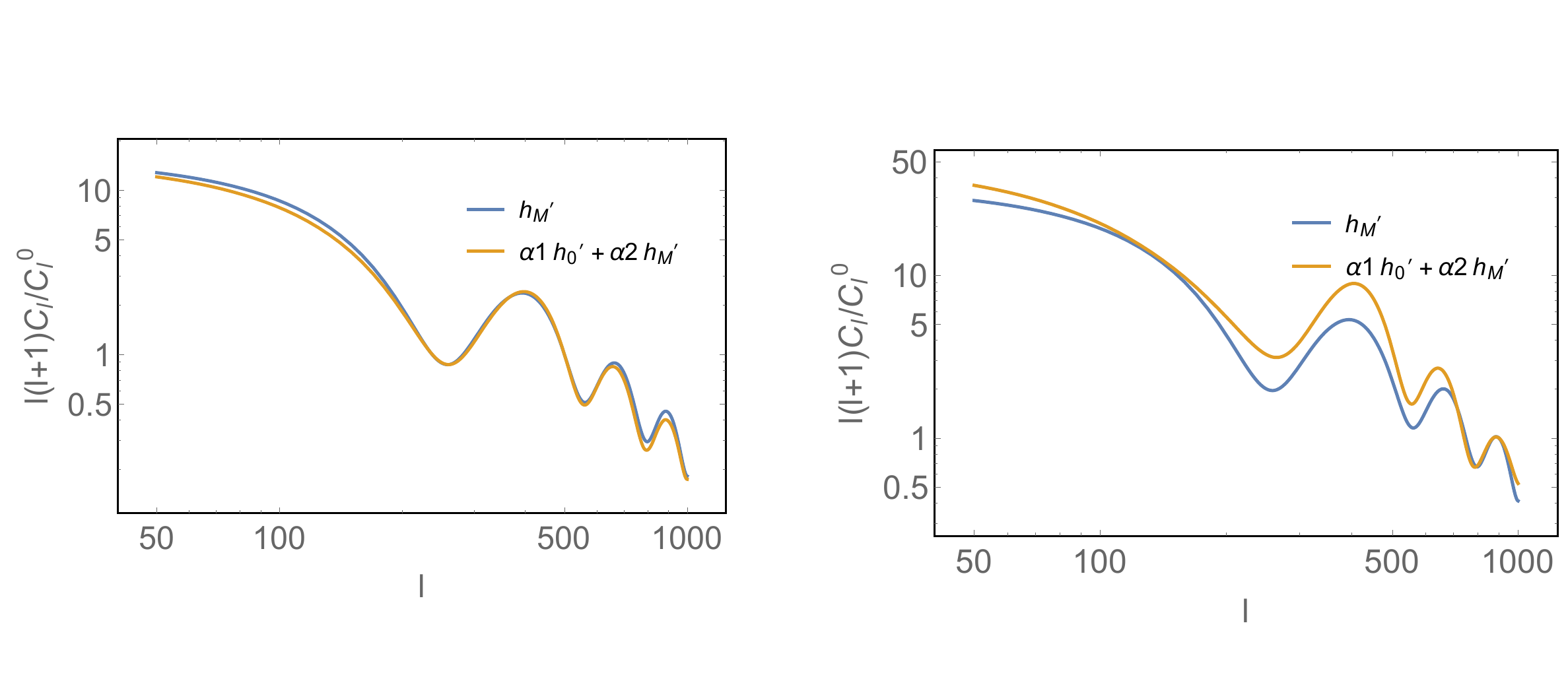}
\caption{The B-mode spectrum $C_{ll}^B$ for  a massive stable graviton and a linear combination of a massive stable graviton and a massless unstable graviton, i.e. a massless graviton with a tachyonic instability in the radiation era.  The mass is taken to be $m=1000 H_0$ and  on the right  $M_{12}^2/ M^2_{22}=1/10$ whilst on the left  $M_{12}^2/M^2_{22}=1/2$. The tensor to scalar ratio is set to $10^{-2}$. Notice that the plateau at low $l$ is hardly modified by the unstable mode whilst the first few peaks are shifted compared to a purely massive graviton.  }
\label{CMB_instab}
\end{center}
\end{figure}
This result is a generalisation of (\ref{Diagonalised:couplings}) in radiation domination.  It can be extrapolated to the transition region by redefining
\be
\tilde M^2_{ij}=M^2_{ij} + \frac{\beta_i\beta_j}{\beta_1^2+\beta_2^2} 3\omega H^2
\ee
where $\omega=0$ in the matter era and $\omega=1/3$ in the radiation era.

During radiation domination the instability dominates and we can approximate
\bea
\tilde M_{22}^2\approx\frac{\beta_2^2}{\tau^2},\hspace{1cm}\tilde M_{12}^2\approx\frac{\beta_1\beta_2}{\tau^2}
\eea
so we have that $\tilde M_{12}^2/\tilde M^2_{22}\approx\beta_1/\beta_2$. Then (\ref{InstabilityDiagonalised:couplings}) reduces to
\bea
\delta S_{\rm{matter}}=\frac{1}{2M_{\rm Pl}}\int \d^4 xa^2\frac{f^m_{ij}}{a}T^{ij},
\eea
which implies that during radiation domination there is no coupling of the unstable mode to matter. To analyse how this changes the power spectrum of B-modes  we set $\beta_2=\beta_1=1$ and we choose a large mass for the graviton of $1000\ H_0$, i.e. leading to a plateau at low $l$. We also vary the ratio $M_{12}^2/ M^2_{22}$. We find that the effect of the instability is very mild  only affecting the first few peaks of the B-mode spectrum, see figure \ref{CMB_instab}. We have imposed that the initial conditions for the unstable modes are such that at recombination its magnitude does not exceed the one of the massive graviton. In principle, the
effects of the unstable mode can be reduced by choosing even lower initial values. This choice  guarantees that the mode does not go non-linear before recombination. As a result, the power spectrum even in the presence of a tachyonic instability, here tamed by the initial conditions, is characterised by the typical plateau of massive gravitons at low $l$ and a shifted structure of peaks compared to a purely massive graviton case.
\section{Conclusions}
In this paper we have analysed the effect that a modification of gravity has on the B-mode power spectrum. Our results suggest that if $r$ becomes observable, the constraints on modified gravity theories will improve greatly. In particular we have studied the effect that massive gravity has on the B mode power spectrum. We have found analytical  expressions for the massive tensor modes valid during  matter and radiation domination. With this result we have found that the most important effect  a massive graviton has on the the CMB is  the presence of  a plateau at small $l$, as the source function for gravitational waves is constant outside the horizon.

We have also studied multiple gravitons. Using our analytical results we have  shown how the effects of massive gravity arise in the presence of a combined coupling to gravity. In general the massive graviton is always coupled to matter and its effect cannot be removed by tuning the mass parameters of the models as long as one massless graviton is present.  Moreover we have also included the effects of the tachyonic instability of doubly coupled bigravity that arises in the radiation era. The instability affects  the massless graviton which becomes tachyonic in the radiation era. As this mode  does not couple to matter during radiation domination, its effect is very mild and does not alter the existence of a plateau at low $l$, a feature of massive gravitons. This result is valid as long as the initial solutions of the unstable mode are reduced at the end of inflation \cite{Comelli:2015pua,Brax:2017uwg}.

A future detection of primordial B-modes  would  help improving the bounds on the mass of the graviton. Indeed  measuring the position of the first peaks and finding   a plateau at low $l$ would give clues about the existence
 of one massless graviton mixed with one massive graviton. Hence the results in this paper could help constraining  multigravity, and we intend to perform a  Fisher matrix analysis to find out whether the effects we have described would  be detectable in the next generation of CMB experiments.
\section*{Acknowledgements} SC is supported by  Conicyt through its programme Becas Chile and the Cambridge overseas Trust. SC and ACD  acknowledges partial support from STFC under grants ST/L000385/1 and ST/L000636/1.
This work is supported in part by the EU Horizon 2020 research and innovation programme under the Marie-Sklodowska grant No. 690575. This article is based upon work related to the COST Action CA15117 (CANTATA) supported by COST (European Cooperation in Science and Technology).
\appendix

\section{Matching solutions}
For a massive graviton we have that the solutions are for matter domination
\bea
h=\frac{3}{k\tau} j_0(\frac{1}{3}m\bar H_0^2\tau^3)j_1(k\tau),
\eea
and during radiation domination
\bea
h=(mH_0\tau^2/2)^{1/4}j_{-1/4}(m\hat H_0\tau^2/2)j_0(k\tau).
\eea
Both solutions are constant outside the horizon $k\tau,ma^2\ll 1$, to then start decaying. This makes  the matching with inflation  easy as after leaving the horizon during inflation tensor modes freeze out, and its amplitude are constant.

At matter-radiation equality  we can assume that the transition  was instantaneous and then match the solution and its derivative at $\tau_{eq}$. This leads to,
\bea
h=
\begin{cases}
(m\hat H_0\tau^2/2)^{1/4}j_{-1/4}(m\hat H_0\tau^2/2)j_0(k\tau) & \tau<\tau_{eq}\\
 \frac{3}{k\tau} j_0(\frac{1}{3}m\bar H_0^2\tau^3)\left(Aj_1(k\tau)+By_1(k\tau)\right) & \tau>\tau_{eq}
\end{cases},
\eea
where
\bea
A&=& \frac{\sqrt{\pi } k\tau _{\text{eq}}  m \tau _{\text{eq}} \csc \left(\frac{m \tau
   _{\text{eq}}^3}{3}\right)}{18\ 2^{3/4} k^2} \left(8 \, _0\tilde{F}_1\left(;\frac{1}{4};-\frac{1}{16}
   m^2 \tau _{\text{eq}}^4\right) \sin \left(k \tau _{\text{eq}}\right) \left(k \tau
   _{\text{eq}} \sin \left(k \tau _{\text{eq}}\right)+\cos \left(k \tau
   _{\text{eq}}\right)\right)+   \right.\nonumber\\
 & & \left. \, _0\tilde{F}_1\left(;\frac{5}{4};-\frac{1}{16} m^2 \tau
   _{\text{eq}}^4\right) \left(-2 m \tau _{\text{eq}}^3 \sin \left(k \tau
   _{\text{eq}}\right) \cot \left(\frac{m \tau _{\text{eq}}^3}{3}\right) \left(k \tau
   _{\text{eq}} \sin \left(k \tau _{\text{eq}}\right)+\cos \left(k \tau
   _{\text{eq}}\right)\right)+  \right. \right.\nonumber\\
 & & \left.\left. 4 k \tau _{\text{eq}}+3 \sin \left(2 k \tau
   _{\text{eq}}\right)-2 k \tau _{\text{eq}} \cos \left(2 k \tau
   _{\text{eq}}\right)\right)\right),\\
   B&=& \frac{\sqrt{\pi } k\tau _{\text{eq}}  m \tau _{\text{eq}} \csc \left(\frac{m \tau
   _{\text{eq}}^3}{3}\right)}{18\ 2^{3/4} k^2} \left(\, _0\tilde{F}_1\left(;\frac{5}{4};-\frac{1}{16} m^2 \tau _{\text{eq}}^4\right)
   \left(-2 k^2 \tau _{\text{eq}}^2+m \tau _{\text{eq}}^3 \cot \left(\frac{m \tau
   _{\text{eq}}^3}{3}\right) \left(k \tau _{\text{eq}} \sin \left(2 k \tau
   _{\text{eq}}\right)+\right.\right. \right.\nonumber\\
 & &\left. \left.\left.\cos \left(2 k \tau _{\text{eq}}\right)-1\right)-2 k \tau
   _{\text{eq}} \sin \left(2 k \tau _{\text{eq}}\right)-3 \cos \left(2 k \tau
   _{\text{eq}}\right)+3\right)-\right.\nonumber\\
 & & 4 \left.\, _0\tilde{F}_1\left(;\frac{1}{4};-\frac{1}{16} m^2
   \tau _{\text{eq}}^4\right) \left(k \tau _{\text{eq}} \sin \left(2 k \tau
   _{\text{eq}}\right)+\cos \left(2 k \tau _{\text{eq}}\right)-1\right)\right).
\eea
It can also be assumed that the transition is much smaller than the wavelenght and use the WKB approximation. This approach was followed by~\cite{Pritchard:2004qp} for the case of massless gravity. Its application to our case its straightforward, but is beyond	 the reach or our paper.

\providecommand{\href}[2]{#2}\begingroup\raggedright\endgroup

\begin{thebibliography}{99}
\bibitem{Suzuki:2015zzg}
  A.~Suzuki {\it et al.} [POLARBEAR Collaboration],
  ``The POLARBEAR-2 and the Simons Array Experiment,''
  J.\ Low.\ Temp.\ Phys.\  {\bf 184} (2016) no.3-4,  805
  doi:10.1007/s10909-015-1425-4
  [arXiv:1512.07299 [astro-ph.IM]].
\bibitem{Watts:2015eqa}
  D.~J.~Watts {\it et al.},
  ``Measuring the Largest Angular Scale CMB B-mode Polarization with Galactic Foregrounds on a Cut Sky,''
  Astrophys.\ J.\  {\bf 814} (2015) no.2,  103
  doi:10.1088/0004-637X/814/2/103
  [arXiv:1508.00017 [astro-ph.CO]].
\bibitem{Abazajian:2016yjj}
  K.~N.~Abazajian {\it et al.} [CMB-S4 Collaboration],
  ``CMB-S4 Science Book, First Edition,''
  arXiv:1610.02743 [astro-ph.CO].
\bibitem{Kamionkowski:1996zd}
  M.~Kamionkowski, A.~Kosowsky and A.~Stebbins,
  ``A Probe of primordial gravity waves and vorticity,''
  Phys.\ Rev.\ Lett.\  {\bf 78} (1997) 2058
  doi:10.1103/PhysRevLett.78.2058
  [astro-ph/9609132].
  \bibitem{Zaldarriaga:1996xe}
  M.~Zaldarriaga and U.~Seljak,
  ``An all sky analysis of polarization in the microwave background,''
  Phys.\ Rev.\ D {\bf 55} (1997) 1830
  doi:10.1103/PhysRevD.55.1830
  [astro-ph/9609170].
\bibitem{Koyama:2015vza}
  K.~Koyama,
  ``Cosmological Tests of Modified Gravity,''
  Rept.\ Prog.\ Phys.\  {\bf 79} (2016) no.4,  046902
  doi:10.1088/0034-4885/79/4/046902
  [arXiv:1504.04623 [astro-ph.CO]].
\bibitem{Clifton:2011jh}
  T.~Clifton, P.~G.~Ferreira, A.~Padilla and C.~Skordis,
  ``Modified Gravity and Cosmology,''
  Phys.\ Rept.\  {\bf 513} (2012) 1
  doi:10.1016/j.physrep.2012.01.001
  [arXiv:1106.2476 [astro-ph.CO]].
\bibitem{Joyce:2014kja}
  A.~Joyce, B.~Jain, J.~Khoury and M.~Trodden,
  ``Beyond the Cosmological Standard Model,''
  Phys.\ Rept.\  {\bf 568} (2015) 1
  doi:10.1016/j.physrep.2014.12.002
  [arXiv:1407.0059 [astro-ph.CO]].
\bibitem{Jimenez:2015bwa}
  J.~Beltran Jimenez, F.~Piazza and H.~Velten,
  ``Evading the Vainshtein Mechanism with Anomalous Gravitational Wave Speed: Constraints on Modified Gravity from Binary Pulsars,''
  Phys.\ Rev.\ Lett.\  {\bf 116} (2016) no.6,  061101
  doi:10.1103/PhysRevLett.116.061101
  [arXiv:1507.05047 [gr-qc]].

\bibitem{ArkaniHamed:2002sp}
  N.~Arkani-Hamed, H.~Georgi and M.~D.~Schwartz,
  ``Effective field theory for massive gravitons and gravity in theory space,''
  Annals Phys.\  {\bf 305} (2003) 96
  doi:10.1016/S0003-4916(03)00068-X
  [hep-th/0210184].
\bibitem{deRham:2010ik}
  C.~de Rham and G.~Gabadadze,
  ``Generalization of the Fierz-Pauli Action,''
  Phys.\ Rev.\ D {\bf 82} (2010) 044020
  doi:10.1103/PhysRevD.82.044020
  [arXiv:1007.0443 [hep-th]].
\bibitem{deRham:2010kj}
  C.~de Rham, G.~Gabadadze and A.~J.~Tolley,
  ``Resummation of Massive Gravity,''
  Phys.\ Rev.\ Lett.\  {\bf 106}, 231101 (2011)
  doi:10.1103/PhysRevLett.106.231101
  [arXiv:1011.1232 [hep-th]].
\bibitem{Hassan:2011hr}
  S.~F.~Hassan and R.~A.~Rosen,
  ``Resolving the Ghost Problem in non-Linear Massive Gravity,''
  Phys.\ Rev.\ Lett.\  {\bf 108}, 041101 (2012)
  doi:10.1103/PhysRevLett.108.041101
  [arXiv:1106.3344 [hep-th]].
\bibitem{Hassan:2011zd}
  S.~F.~Hassan and R.~A.~Rosen,
  ``Bimetric Gravity from Ghost-free Massive Gravity,''
  JHEP {\bf 1202}, 126 (2012)
  doi:10.1007/JHEP02(2012)126
  [arXiv:1109.3515 [hep-th]].
\bibitem{DAmico:2011eto}
  G.~D'Amico, C.~de Rham, S.~Dubovsky, G.~Gabadadze, D.~Pirtskhalava and A.~J.~Tolley,
``Massive Cosmologies,''
  Phys.\ Rev.\ D {\bf 84} (2011) 124046
  doi:10.1103/PhysRevD.84.124046
  [arXiv:1108.5231 [hep-th]].
\bibitem{Enander:2014xga}
  J.~Enander, A.~R.~Solomon, Y.~Akrami and E.~Mortsell,
  ``Cosmic expansion histories in massive bigravity with symmetric matter coupling,''
  JCAP {\bf 1501} (2015) 006
  doi:10.1088/1475-7516/2015/01/006
  [arXiv:1409.2860 [astro-ph.CO]].
\bibitem{Gumrukcuoglu:2015nua}
  A.~E.~Gumrukcuoglu, L.~Heisenberg, S.~Mukohyama and N.~Tanahashi,
  ``Cosmology in bimetric theory with an effective composite coupling to matter,''
  JCAP {\bf 1504} (2015) no.04,  008
  doi:10.1088/1475-7516/2015/04/008
  [arXiv:1501.02790 [hep-th]].
\bibitem{Lagos:2015sya}
  M.~Lagos and J.~Noller,
 ``New massive bigravity cosmologies with double matter coupling,''
  JCAP {\bf 1601} (2016) no.01,  023
  doi:10.1088/1475-7516/2016/01/023
  [arXiv:1508.05864 [gr-qc]].
\bibitem{Deffayet:2009mn}
  C.~Deffayet, S.~Deser and G.~Esposito-Farese,
  ``Generalized Galileons: All scalar models whose curved background extensions maintain second-order field equations and stress-tensors,''
  Phys.\ Rev.\ D {\bf 80} (2009) 064015
  doi:10.1103/PhysRevD.80.064015
  [arXiv:0906.1967 [gr-qc]].
\bibitem{Lombriser:2015sxa}
  L.~Lombriser and A.~Taylor,
  ``Breaking a Dark Degeneracy with Gravitational Waves,''
  JCAP {\bf 1603} (2016) no.03,  031
  doi:10.1088/1475-7516/2016/03/031
  [arXiv:1509.08458 [astro-ph.CO]].
\bibitem{Brax:2015dma}
  P.~Brax, C.~Burrage and A.~C.~Davis,
  ``The Speed of Galileon Gravity,''
  JCAP {\bf 1603} (2016) no.03,  004
  doi:10.1088/1475-7516/2016/03/004
  [arXiv:1510.03701 [gr-qc]].
\bibitem{Bettoni:2016mij}
  D.~Bettoni, J.~M.~Ezquiaga, K.~Hinterbichler and M.~Zumalacárregui,
  ``Speed of Gravitational Waves and the Fate of Scalar-Tensor Gravity,''
  Phys.\ Rev.\ D {\bf 95}, no. 8, 084029 (2017)
  doi:10.1103/PhysRevD.95.084029
  [arXiv:1608.01982 [gr-qc]];

\bibitem{TheLIGOScientific:2017qsa}
  B.~P.~Abbott {\it et al.} [LIGO Scientific and Virgo Collaborations],
  ``GW170817: Observation of Gravitational Waves from a Binary Neutron Star Inspiral,''
  Phys.\ Rev.\ Lett.\  {\bf 119} (2017) no.16,  161101
  doi:10.1103/PhysRevLett.119.161101
  [arXiv:1710.05832 [gr-qc]].
 \bibitem{GWHordensky}
  P.~Creminelli and F.~Vernizzi,
  ``Dark Energy after GW170817,''
  arXiv:1710.05877 [astro-ph.CO];
    J.~M.~Ezquiaga and M.~Zumalacárregui,
  `Dark Energy after GW170817,''
  arXiv:1710.05901 [astro-ph.CO];
   J.~Sakstein and B.~Jain,
  ``Implications of the Neutron Star Merger GW170817 for Cosmological Scalar-Tensor Theories,''
  arXiv:1710.05893 [astro-ph.CO];
    T.~Baker, E.~Bellini, P.~G.~Ferreira, M.~Lagos, J.~Noller and I.~Sawicki,
  ``Strong constraints on cosmological gravity from GW170817 and GRB 170817A,''
  arXiv:1710.06394 [astro-ph.CO].
\bibitem{Burrage:2016myt}
  C.~Burrage, S.~Cespedes and A.~C.~Davis,
  ``Disformal transformations on the CMB,''
  JCAP {\bf 1608}, no. 08, 024 (2016)
  doi:10.1088/1475-7516/2016/08/024
  [arXiv:1604.08038 [gr-qc]].
\bibitem{Raveri:2014eea}
  M.~Raveri, C.~Baccigalupi, A.~Silvestri and S.~Y.~Zhou,
  ``Measuring the speed of cosmological gravitational waves,''
  Phys.\ Rev.\ D {\bf 91} (2015) no.6,  061501
  doi:10.1103/PhysRevD.91.061501
  [arXiv:1405.7974 [astro-ph.CO]].
\bibitem{Cornish:2017jml}
  N.~Cornish, D.~Blas and G.~Nardini,
  ``Bounding the speed of gravity with gravitational wave observations,''
  arXiv:1707.06101 [gr-qc].
\bibitem{deRham:2016nuf}
  C.~de Rham, J.~T.~Deskins, A.~J.~Tolley and S.~Y.~Zhou,
  ``Graviton Mass Bounds,''
  Rev.\ Mod.\ Phys.\  {\bf 89} (2017) no.2,  025004
  doi:10.1103/RevModPhys.89.025004
  [arXiv:1606.08462 [astro-ph.CO]].
\bibitem{Pritchard:2004qp}
  J.~R.~Pritchard and M.~Kamionkowski,
  ``Cosmic microwave background fluctuations from gravitational waves: An Analytic approach,''
  Annals Phys.\  {\bf 318} (2005) 2
  doi:10.1016/j.aop.2005.03.005
  [astro-ph/0412581].
\bibitem{Fasiello:2015csa}
  M.~Fasiello and R.~H.~Ribeiro,
  ``Mild bounds on bigravity from primordial gravitational waves,''
  JCAP {\bf 1507} (2015) no.07,  027
  doi:10.1088/1475-7516/2015/07/027
  [arXiv:1505.00404 [astro-ph.CO]].

\bibitem{Dubovsky:2009xk}
  S.~Dubovsky, R.~Flauger, A.~Starobinsky and I.~Tkachev,
  ``Signatures of a Graviton Mass in the Cosmic Microwave Background,''
  Phys.\ Rev.\ D {\bf 81}, 023523 (2010)
  doi:10.1103/PhysRevD.81.023523
  [arXiv:0907.1658 [astro-ph.CO]].
\bibitem{Brax:2016ssf}
  P.~Brax, A.~C.~Davis and J.~Noller,
  ``Dark Energy and Doubly Coupled Bigravity,''
  Class.\ Quant.\ Grav.\  {\bf 34} (2017) no.9,  095014
  doi:10.1088/1361-6382/aa6856
  [arXiv:1606.05590 [gr-qc]].
\bibitem{Comelli:2015pua}
  D.~Comelli, M.~Crisostomi, K.~Koyama, L.~Pilo and G.~Tasinato,
  ``Cosmology of bigravity with doubly coupled matter,''
  JCAP {\bf 1504} (2015) 026
  doi:10.1088/1475-7516/2015/04/026
  [arXiv:1501.00864 [hep-th]].
\bibitem{Brax:2017uwg}
  P.~Brax and P.~Valageas,
  arXiv:1712.04520 [gr-qc].
\bibitem{Bernard:2015mkk}
  L.~Bernard, C.~Deffayet and M.~von Strauss,
  JCAP {\bf 1506} (2015) 038
  doi:10.1088/1475-7516/2015/06/038
  [arXiv:1504.04382 [hep-th]].  
%

%
%
%
%
%
%
%
%
%
%
%
%
%
%
%
%
%
%
%
%
%
%
%
%
%
%
%
%
%

\end{thebibliography}
\end{document}